\documentclass{aa}
\usepackage{graphicx,txfonts,float,footmisc,twoopt}
\usepackage{svg}
\usepackage{natbib} 
\usepackage{xcolor}
\usepackage[normalem]{ulem}
\usepackage{amssymb,amsmath}
\usepackage{gensymb}
\usepackage[version=3]{mhchem}
\usepackage[]{hyperref}

\begin{document} 

%### Title
%%%%%%%%%%%%%%%%%%%%%%%%%%%%%%%%%%%%%%%%
\title{SPH modelling of AGB wind morphology in hierarchical triple systems \& comparison to observation of R Aql}

\author{J. Malfait \inst{1} \and  L. Siess \inst{2} \and O. Vermeulen \inst{1} \and M. Esseldeurs \inst{1} \and S. H. J. Wallstr\"om \inst{1} \and A. M. S. Richards \inst{3} \and F. De Ceuster \inst{1} \and S. Maes \inst{1} \and J. Bolte \inst{4} \and L. Decin \inst{1}}

\institute{ Institute of Astronomy, KU Leuven, Celestijnenlaan 200D, 3001 Leuven, Belgium \and Institut d'Astronomie et d'Astrophysique, Universit\'e Libre de Bruxelles (ULB), CP 226, 1050 Brussels, Belgium \and Jodrell Bank Centre for Astrophysics, Department of Physics and Astronomy, University of Manchester, Oxford Road, Manchester M13 9PL, UK \and Department of Mathematics, Kiel University, Heinrich-Hecht-Platz 6, 24118 Kiel, Germany
}

\date{Accepted }

\abstract
% context heading (optional)
% {} leave it empty if necessary  
{Complex asymmetric 3D structures are observed in the outflows of evolved low- and intermediate-mass stars, and are believed to be shaped through the interaction of companions that remain hidden within the dense wind. One example is the AGB star R Aql, for which ALMA observations reveal complex wind structures, that might originate from a higher-order stellar system.}
% aims heading (mandatory)
{We investigate how triple systems can shape the outflow of Asymptotic Giant Branch (AGB) stars, and characterise the different wind structures that form. For simplicity, we solely focus on coplanar systems in a hierarchical, stable orbit, consisting of an AGB star with one relatively close companion, and another one at large orbital separation.}
% methods heading (mandatory)	
{We model a grid of hierarchical triple systems including a wind-launching AGB star, with the smoothed-particle-hydrodynamic \textsc{Phantom} code. We vary the outer companion mass, the AGB wind velocity and the orbital eccentricities to study the impact of these parameters on the wind morphology. To study the impact of adding a triple companion, we additionally model and analyse a small grid of binary sub-systems, for comparison. To investigate if R Aql could be shaped by a triple system, we post-process one of our triple models with a radiative transfer routine, and compare this to data of the ALMA ATOMIUM programme.}
% results heading (mandatory)
{The characteristic wind structures resulting from a hierarchical triple system are the following. A large two-edged spiral wake results behind the outer companion star. This structure lies on top of the spiral structure formed by the close binary, which is itself affected by the orbital motion around the system centre-of-mass, such that it resembles a snail-shell pattern. This dense inner spiral pattern interacts with, and strongly impacts, the spiral wake of the outer companion, resulting in a waved-pattern on the outer edge of this spiral wake. The higher the mass of the outer companion, the larger the density enhancement, and the more radially compressed the outer spiral. Lowering the wind velocity has a similar effect, and additionally results in an elongation of the global wind morphology. 
Introducing eccentricity in the inner and outer orbit of the hierarchical system, results in complex phase-dependent wind-companion interactions, and consequently in asymmetries in the inner part of the wind and the global morphology, respectively. 
From the comparison of our models to the observations of R Aql, we conclude that this circumstellar environment might be shaped by a similar system as the ones modelled in this work, but an elaborate study of the observational data is needed to determine better the orbital parameters and characteristics of the central system.}
% conclusions heading (optional), leave it empty if necessary 
{The modelled outflow of an AGB star in a coplanar hierarchical systems is characterised by a large-scale spiral wake with a waved outer edge, attached to the outer companion, on top of a compact inner spiral pattern that resembles a snail-shell pattern.}

\keywords{Stars: AGB -- Stars: winds, outflows -- Hydrodynamics -- binaries -- Methods: numerical}

\authorrunning{Malfait et al.}
\maketitle

\section{Introduction}
\label{ch:intro}

At the end of the core-He burning phase, evolved low- and intermediate-mass stars enter the Asymptotic Giant Branch (AGB) phase. During this phase, the stars expand and cool down, and lose their outer envelopes through pulsation-enhanced dust-driven winds. Afterwards, they evolve through the post-AGB and planetary nebulae (PNe) evolutionary phases, to become white dwarfs.
Many of these AGB successors host binary or higher-order (more than two stars) stellar systems \citep[see e.g.][]{DeMarco2013,Oomen2018,Escorza2023}, and have very complex morphologies that are believed to be related to this multiple nature \citep{Jones2017,Akashi2017,Demarco2009,Garcia2018,Corporaal2023,Verhamme2024}. 
Similarly, AGB outflows are recently being observed with high angular and spectral resolution, revealing non-spherically symmetric circumstellar environments in which complex structures such as arcs, spirals, disks, and bipolar outflows are discovered, with the main shaping mechanism of these complexities believed to be binarity \citep{Mauron2006,Maercker2012,Decin2020,Danilovich2024}. 
Moreover, both in PNe and in AGB outflows, the complexity of some observed morphological structures suggests that they may be created by the impact of more than one companion. An example is the AGB star R Aql, which is observed within the ATOMIUM ALMA large programme, and has a circumstellar envelope (CSE) with a lot of intricate spiral- and arc-like structures \citep{Decin2020}.
Recent theoretical and observational studies strengthen the hypothesis that several PNe are shaped by triple stellar systems \citep{Akashi2017,Jones2019,Schreier2019,Miszalski2019}.
According to population synthesis studies, $\sim 50\%$ of low-mass stellar systems have at least two, and $13\%$ have three or more stellar components \citep{Offner2023,Tokovinin2014,Gao2022}.
Triple systems often tend to be in a stable hierarchical configuration, meaning that they consist of an inner binary and a distant third component \citep{HUT1983,vandenBerk2007,Tokovinin2021}. 
Although in a hierarchical setup, many triple systems have very complex orbital mechanics, with for example inclined and eccentric orbits in which Kozai-Lidov cycles are at play \citep{Lidov1962,Kozai1962,Mardling2001,Naoz2016}. 
In this work, we take a first step in understanding the effect of a triple stellar system on the structure formation of AGB outflows. We investigate this with the use of smoothed-particle hydrodynamic (SPH) models of triple systems including a wind-launching AGB star.
For this first study, we limit ourselves to the most elementary triple systems that are co-planar and in stable hierarchical orbits, such that complex mechanisms like unstable N-body orbits and Kozai-Lidov cycles are not relevant.

This paper builds on previous studies in which the wind morphology in various binary AGB systems has been investigated in detail with the use of different hydrodynamic models, including \citet{Kim2017,Liu2017,Chen2017,Saladino2018,ElMellah2020,Maes2021,Malfait2021,Esseldeurs2023,Malfait2024}.
We introduce the input physics of our SPH model, and the setup of our grid, consisting of triple and binary systems, in Sect.~\ref{ch:SimulationsMethod}. We first investigate the wind morphology in our set of binary systems in Sect.~\ref{ch:binSim}, before analysing the impact of adding a third body to these simulations in Sect.~\ref{ch:triples}. This analysis includes a study of the impact of varying the wind velocity and orbital eccentricity on the wind morphology. The source of inspiration to study the morphology of triple systems is the AGB star R Aql. We therefore present a short study of ALMA observations of R Aql and compare these to a synthetic simulation of one of our triple hydrodynamic simulations in Sect.~\ref{ch:RAqlSection}. Finally, our conclusions are provided in Sect.~\ref{ch:conclusions}.

\section{Method}
\label{ch:SimulationsMethod}
\subsection{Input physics SPH model}
\label{ch:inputPhysics}

We perform hydrodynamic simulations of an AGB outflow in a hierarchical triple system with the smoothed-particle hydrodynamics (SPH) code \textsc{Phantom} \citep{phantom}. We use the same numerical input physics as described in \cite{Malfait2024}, where more details can be found. 
The input files and output dumps to reproduce the Phantom simulations and figures in this paper are available online (\url{https://zenodo.org/records/12795466}).
In short, the AGB star and companions are modelled as gravity-only sink particles, that can accrete the mass and momentum of gas particles that fall within $0.8$ to $1$ times the setup accretion radius \citep[see][]{phantom,Malfait2024}. 
In wind simulations, SPH particles are launched with an initial velocity $v_{\rm ini}$, on concentric spheres around the primary AGB star \citep{Siess2022}. They are accelerated by a pressure gradient and  with the use of the free-wind approximation, in which the gravity of the AGB star is neglected by the wind particles, as such artificially balancing the gravity by the radiation force \citep{TheunsI,Esseldeurs2023}. 
H\,{\sc i} cooling, following \cite{Spitzer1978}, is included in the energy equation, as explained in \cite{Malfait2024}.
We adopt constant values of the polytropic index $\gamma = 1.2$ and the mean molecular weight $\mu = 2.38$, modelling a molecular wind with a typical, radially decreasing, power-law temperature profile for AGB winds, that strongly depends on the heating and cooling mechanisms included in the energy equation \citep{Millar2004,Maes2021}.

\subsection{Grid setup triples}

\begin{figure}
        \centering
	\includegraphics[width = 0.4 \textwidth]{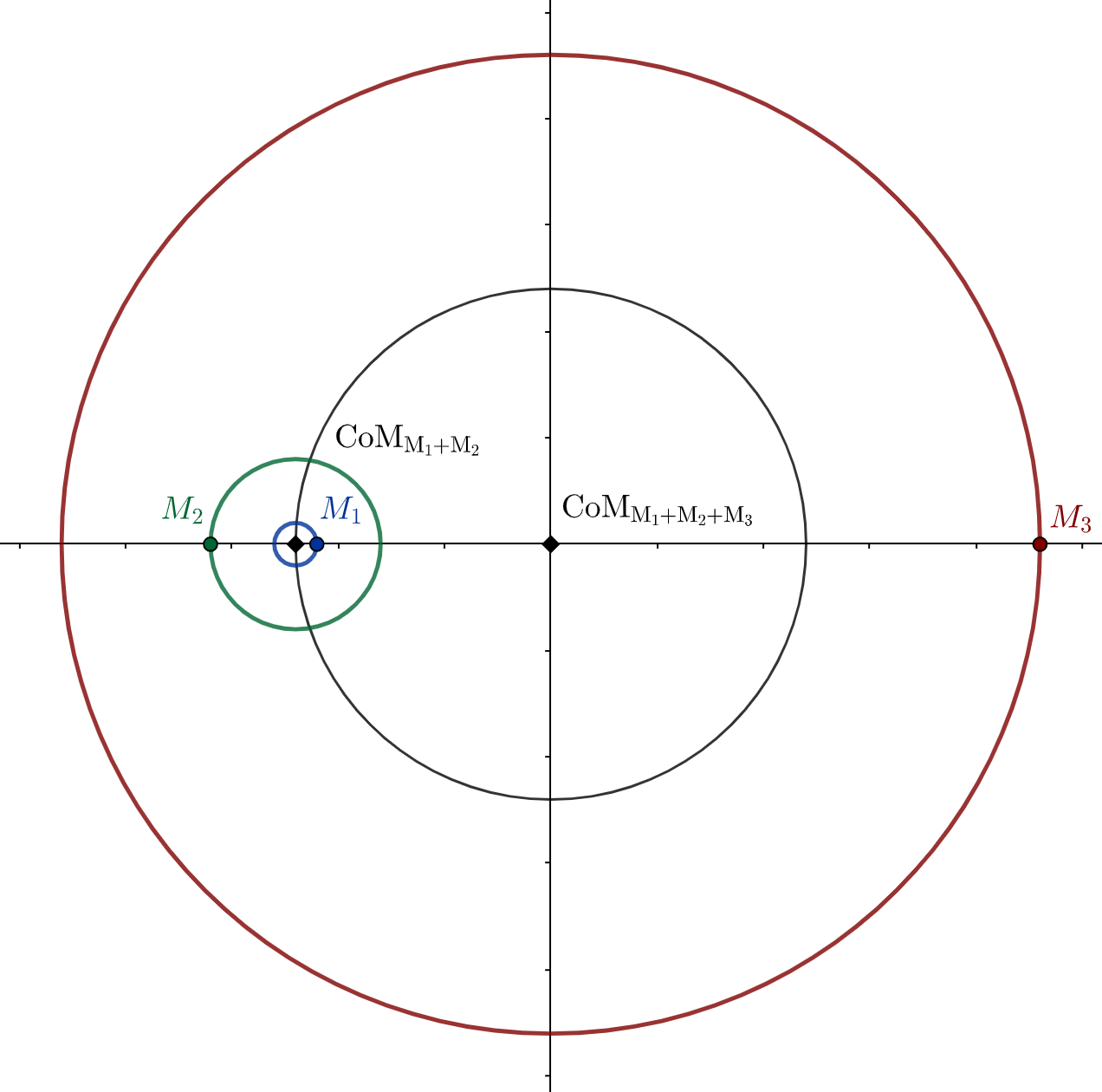}
	\caption{Setup of hierarchical triple with inner binary $M_{\rm 1}$--$M_{\rm 2}$ orbiting around ${\rm{CoM}}_{M_{\rm 1}+M_{\rm 2}}$, and outer companion $M_3$ orbiting around the system's centre-of-mass ${\rm{CoM}}_{M_{\rm 1}+M_{\rm 2}+M_{\rm 3}}$. $M_3$ induces an orbital motion of ${\rm{CoM}}_{M_{\rm 1}+M_{\rm 2}}$, and thereby also $M_1$ and $M_2$, around the system's centre-of-mass.}
	\label{Fig:hierarchTripleSetup}
\end{figure}

Fig.~\ref{Fig:hierarchTripleSetup} illustrates the hierarchical setup used in our models.
It consists of an inner binary given by $M_{\rm 1}$--$M_{\rm 2}$, and a third companion $M_3$. 
The third star and the CoM of the inner binary (${\rm{CoM}}_{M_{\rm 1}+M_{\rm 2}}$) form the outer binary, orbiting the CoM of the entire system (${\rm{CoM}}_{M_{\rm 1}+M_{\rm 2}+M_{\rm 3}}$).
Table~\ref{Ta:setupTableTriples} provides the setup parameters that are the same for all models. The inner binary $M_{\rm 1}$--$M_{\rm 2}$ is set up with a mass ratio $q_1 = M_2/M_1 = 0.25$ in which $M_1 = 1.6 \, {\rm M_{\odot}}$ is the mass of the AGB star, and $M_2 = 0.4 \, {\rm M_{\odot}}$ is the mass of the close companion. The primary has a mass-loss rate of $\dot{M} = 1.1 \times 10^{-6} \, {\rm{M_\odot\,yr^{-1}}}$ and surface temperature $T_1 = 2800 \, {\rm K}$.
These parameters are chosen to correspond to those of the AGB star R Aql, for which high-resolution observations reveal a very intricate wind morphology that might be created by the presence of multiple companions \citep[][see also Sect.~\ref{ch:ObservRAql}]{Decin2020}.

Table \ref{Ta:inputTableTriples} summarises the properties of all our triple simulations.
We start by varying the initial wind velocity $v_{\rm ini}$ between $8 \, {\rm{and}} \, 15 \, {\rm{km\,s^{-1}}}$, as this parameter is known to strongly impact the outflow morphology \citep{Maes2021,Malfait2021,Malfait2024}. Secondly, we vary the mass of the third companion ($M_3$) to $0.1, 0.6\, {\rm{and}} \, 1.2 \,{\rm M_{\odot}}$ to analyse its gravitational influence on the wind. Finally, we investigate the impact of eccentricity on the morphologies in triple systems, by introducing an orbital eccentricity of $e_{\rm i} = 0.4$ and $e_{\rm o} = 0.4$ once in the inner and the outer orbit, respectively. We do this for the triple system v15m06 ($v_{\rm ini} = 15 \, {\rm km \, s^{-1}}$, $M_3 = 0.6 \, {\rm M_\odot}$), which will be used as a reference model for comparison. We select this model, because it has a rather regular, not too complex morphology due to the moderate tertiary mass and high wind velocity. 

The orbital period of the inner binary $M_{\rm 1}$--$M_{\rm 2}$ is $P_{\rm i} = 7.91 \, {\rm yrs}$, and the outer binary orbital period is $P_{\rm o} = 143.0, 128.5 \, {\rm and}\, 115.8 \, {\rm yrs}$ for masses $M_3 = 0.1, 0.6 \,{\rm{and}} \, 1.2 \,{\rm M_{\odot}}$, respectively. 
All systems are modelled during at least $4$ orbital periods of the outer binary. The eccentric systems run until they reach the fourth apastron passage, which occurs at different times in the simulation depending on the initial configuration.

To test if the systems are expected to be dynamically stable, we adopt the commonly used stability criterion of \cite{Mardling2001} for co-planar hierarchical triple systems. This criterion predicts a system to be stable if the ratio of the semi-major axes of the outer to the inner orbit ($a_{\rm{o}}/a_{\rm{i}}$) is larger than the critical ratio $(a_{\rm{o}}/a_{\rm{i}})\vert_{\rm{crit}}$, 
\begin{equation}
	\label{Eq:StabCrit}
	\left. \frac{a_{\rm{o}}}{a_{\rm{i}}}\right \vert_{\rm{crit}} = \frac{2.8}{1-e_{\rm{o}}} \left( \frac{(1 + q_{\rm{o}})\cdot(1 + e_{\rm{o}})}{\sqrt{1-e_{\rm{o}}}} \right)^{2/5} 
\end{equation}
in which the mass ratio of the outer orbit is defined as $q_{\rm{o}}= M_3 /(M_1+M_2)$.
This critical ratio is given in Table~\ref{Ta:inputTableTriples}, and is lower than the ratio $a_{\rm{o}}/a_{\rm{i}} = 7.0$ for all modelled systems, indicating that the systems are expected to be dynamically stable.
As an additional test, the orbit of all systems is integrated with the REBOUND N-body code \citep{rebound} over $1000$ outer orbital periods $P_{\rm o}$, to confirm the stability of the orbits. Note that this test does not take into account the effect of mass-loss, and transfer of mass and angular momentum to the companions, which might influence the orbits of this system during the AGB phase.

\begin{table}
	\caption{Initial setup hierarchical triple models.}
	\begin{center}
		\begin{tabular}{llc}
			\hline 
			\hline
			Parameter     & Unit       &  Initial value \\
			\hline 
			$M_{\rm 1}$      &   [${\rm M_{\odot}}$] &  $1.6$\\
			$M_{\rm 2}$      &   [${\rm M_{\odot}}$] &  $0.4$ \\
			$R_{\rm 1,accr}$  &  [${\rm R_{\odot}}$] &  $258$  \\
			$R_{{\rm 2,accr}}$ & [${\rm R_{\odot}}$] &  $8.6$  \\
			$\dot{M}$		&	[${\rm{M_\odot\,yr^{-1}}}$] & $1.1 \times 10^{-6}$ \\
			$T_{\rm 1}$		&	[K] &  $2800$ \\
			$a_{\rm{i}}$     &            [au] &  $5$ \\
			$a_{\rm{o}}$      &           [au] &  $35$ \\
			$R_{\rm bound}$   &  [au] &  $1500$  \\
                $t_{\rm max} $ &[$P_{\rm o}$] & $4$ \\
			\hline
		\end{tabular}
	\end{center}
	{\textbf{Notes.} \footnotesize{The setup parameters are the same for all hierarchical triple models: $M_{\rm 1}$, $M_{\rm 2}$, $R_{\rm 1,accr}$, and $R_{\rm 2,accr}$ are the initial masses and accretion radii of the primary (AGB) and secondary (companion) star of the inner binary, $\dot{M}$ is the mass-loss rate of the AGB star, and $T_{\rm 1}$ its effective temperature, $a_{\rm{i}}$ and $a_{\rm{o}}$ are the semi-major axes of the inner and outer binary, and $R_{\rm bound}$ is the boundary radius beyond which gas particles are killed. The simulations are computed over at least $4$ outer orbital periods ($P_{\rm o}$).}}
	\label{Ta:setupTableTriples}
\end{table}

\begin{table*}
	\caption{Hierarchical triple model characteristics.} 
	\begin{center}
		\begin{tabular}{lccccccc}
			\hline
			\hline
			&&\\[-2ex]
			Model name &  $v_{\rm ini}$ [${\rm{km\,s^{-1}}}$] & $M_3$ [${\rm M_{\odot}}$] & $e_{\rm{i}}$& $e_{\rm{o}}$ &   $\left . \frac{a_{\rm{o}}}{a_{\rm{i}}}\right \vert_{\rm{crit}}$ & $\varepsilon_{\rm i}$ & $\varepsilon_{\rm o}$ \\
			&&\\[-2ex]
			\hline
			v08m01   & $8$   &  $0.1$ & $0.00$ &$0.00$ & $2.86$ & $2.36$& $0.12$\\
			v08m06   & $8$  & $0.6$ & $0.00$ & $0.00$  & $3.11$ & $2.36$& $0.40$\\
			v08m12   & $8$  & $1.2$ & $0.00$ & $0.00$  & $3.38$ & $2.36$& $0.60$\\
			v15m01   & $15$  & $0.1$  & $0.00$ & $0.00$ & $2.86$ & $1.10$& $0.06$\\
			v15m06   & $15$  & $0.6$ & $0.00$ & $0.00$ & $3.11$ & $1.10$& $0.20$\\
			v15m12   & $15$   & $1.2$ & $0.00$ & $0.00$& $3.38$ & $1.10$& $0.31$\\
   			v15m06e$_{\rm i}$40   & $15$  & $0.6$  & $0.40$ & $0.00$& $3.11$ & $1.83$& $0.20$\\
			v15m06e$_{\rm o}$40   & $15$ & $0.6$  & $0.00$ & $0.40$ & $6.57$ & $1.10$& $0.34$\\

			\hline
		\end{tabular}
	\end{center}
	{\textbf{Notes.} \footnotesize{Hierarchical triple models with their characteristic input values. The model names are set in such a way that the characteristics can be deduced from it, with `v...' denoting the input wind velocity in ${\rm{km\,s^{-1}}}$, `m...' the mass of the third companion multiplied by a factor $10$ in ${\rm M_{\odot}}$, and `e$_{\rm i}$...' and `e$_{\rm o}$...' the value of the non-zero eccentricity of the inner and outer orbit, respectively, multiplied by a factor $100$. $(a_{\rm{o}}/a_{\rm{i}})\vert_{\rm{crit}}$ gives the critical ratio of the outer to inner semi-major axis below which the system is expected to become unstable according to the stability criterium of \cite{Mardling2001} (Eq.~\ref{Eq:StabCrit}), with $a_{\rm{o}}/a_{\rm{i}} = 35/5 = 7 > (a_{\rm{o}}/a_{\rm{i}})\vert_{\rm{crit}}$ for all models. $\varepsilon_{\rm i}$ and $\varepsilon_{\rm o}$ are the morphological classification parameters for the inner and outer orbit, calculated from Eq.\ref{Eq:varepsilon}, following the definition introduced by \cite{Maes2021}.}}
	\label{Ta:inputTableTriples}
\end{table*}

\subsection{Grid setup binaries}
To analyse the impact of adding a tertiary component to a binary system, we first model a limited set of binary models that are subsystems of the hierarchical triple, as described in Fig.~\ref{Fig:hierarchTripleSetup} and Tables~\ref{Ta:setupTableTriples} and \ref{Ta:inputTableTriples}. 
The setup of the binary models is given in Table~\ref{Ta:inputTableBinaries}.
Firstly, we investigate the wind morphology of the inner binary ($M_{\rm 1}$--$M_{\rm 2}$), for the two different initial wind velocities of $8$ and $15 \, {\rm km \, s^{-1}}$ (models v08bin$_{\rm i}$ and v15bin$_{\rm i}$), to which an outer companion $M_3$ is added in the triple simulations.
Additionally, we model one circular and one eccentric ($e=0.4$) outer binary system ($M_{\rm{p}} = M_{\rm 1}+M_{\rm 2}$)--($M_{\rm 3}$), to study the effect of splitting up the primary into a close binary (models v15bin$_{\rm o}$m06 and v15bin$_{\rm o}$m06e40).

\begin{table*}
	\caption{Binary model characteristics.} 
	\begin{center}
		\begin{tabular}{lccccccc}
			\hline
			\hline
			&&\\[-2ex]
			Model name &  $v_{\rm ini}$ [${\rm{km\,s^{-1}}}$] & $M_{\rm p}$ [${\rm M_{\odot}}$] & $M_{\rm s}$ [${\rm M_{\odot}}$] & $a$ [au] & $e$ & $\varepsilon$ & $R_{\rm{f}}$\\
			&&\\[-2ex]
			\hline
			v08bin$_{\rm i}$   & $8$   &  $1.6$ & $0.4$ &$5$ & $0$ & $2.36$ & $0.68$\\
			v15bin$_{\rm i}$   & $15$   &  $1.6$ & $0.4$ &$5$ & $0$ & $1.10$ & $0.80$\\
                v15bin$_{\rm o}$m06 & $15$ & $2.0$ & $0.6$ & $35$ & $0$ & $0.20$ & $0.89$\\
                v15bin$_{\rm o}$m06e40 & $15$ & $2.0$ & $0.6$ & $35$ & $0.4$ & $0.34$ & \\

			\hline
		\end{tabular}
	\end{center}
	{\textbf{Notes.} \footnotesize{Binary models with their characteristic input values, with $M_{\rm p}$ the mass of the primary, and $M_{\rm s}$ the mass of the secondary star. The model names are set in such a way that the characteristics can be deduced from it, with `v...' denoting the input wind velocity in ${\rm{km\,s^{-1}}}$, `m...' the mass of the companion multiplied by a factor $10$ in ${\rm M_{\odot}}$ in case of the outer orbit, and `e' the value of the non-zero eccentricity multiplied by a factor $100$. The subscript $_{\rm i,o}$ indicates if the binary represents the inner or outer orbit of the hierarchical triple systems that are described in Tables~\ref{Ta:setupTableTriples} and ~\ref{Ta:inputTableTriples}. $\varepsilon$ is the morphological classification parameter calculated from Eq.\ref{Eq:varepsilon}, following the definition introduced by \cite{Maes2021}, and $R_{\rm f}$ is the theoretical flattening ratio from Eq.~\ref{Eq:R_f}, as defined by \cite{Malfait2021}.}}
	\label{Ta:inputTableBinaries}
\end{table*}

\subsection{Classification parameter and flattening ratio}
\label{ch:method:EpsFlR}
The strength of the wind-companion interaction and the expected deviation from a spherically symmetric morphology can be estimated with the use of a classification parameter $\varepsilon$, as proposed by \cite{Maes2021}, and applied by \cite{Malfait2021}. In a binary system, $\varepsilon$ is defined as the ratio of gravitational energy density around the companion to the kinetic energy of the wind, 
\begin{equation}
    \varepsilon = \frac{e_{\rm grav}}{e_{\rm kin}} = \frac{(24 \,G^3 M_{\rm s}^2 M_{\rm p})^{1/3}}{v_{\rm w}^2 \, a \, (1-e)}
    \label{Eq:varepsilon}
\end{equation}
with $v_{\rm w}$ an estimate of the wind velocity at the location of the companion. This velocity is approximated by the vector sum of the wind velocity at the orbital separation of the companion in a 1D (spherically symmetric) simulation without companion, and the average orbital velocity of the primary star, to account for the effect of the orbital motion on the wind velocity field. As such, $v_{\rm w}$ is 
\begin{equation}
    v_{\rm w} = \sqrt{v_{\rm 1D}(r = a)^2 + v_{\rm orb, p}^2} .
\end{equation}
We anticipate that when $\varepsilon \ll 1$, a quasi-spherical outflow with a thin, low-density contrast spiral feature will result; for $\varepsilon \sim 1$, a regular Archimedes spiral will form; and for $\varepsilon \gg 1$, a more complex asymmetrical morphology will be shaped \citep{Maes2021}.
For our triple systems, we calculate $\varepsilon$ on $2$ hierarchical levels, for the inner system $M_1$--$M_2$ ($\varepsilon_{\rm i}$) and for the outer binary system $(M_1 + M_2)$--$M_3$ ($\varepsilon_{\rm o}$). The values are provided in Table~\ref{Ta:inputTableTriples} and are used throughout our analysis.

Further, it is known that the impact of a binary companion results into a flattening, or elongation of the morphology, by the induced orbital motion of AGB star \citep{Maes2021,Malfait2021}. As explained by \cite{Malfait2021}, we theoretically expect that the AGB star launches wind material with a velocity $\vec{v_\text{w}} = \vec{v_\text{ini}}+\vec{v_{\text{orb,p}}}$, the vector sum of the input wind velocity and AGB orbital velocity. Because the orbital velocity vector lies in the orbital plane, the resulting wind has a larger velocity component in the orbital plane direction, which results in an elongation of the morphology. 
As illustrated by \cite{Malfait2021}, this flattening can be quantified by the ratio $R_{\rm f}$ of the height to the width of the edge-on wind structures.
Theoretically, when neglecting the impact of the direct gravitational attraction of matter by the companion on the shape of the arc-structures, this flattening ratio is expected to be
\begin{equation}
    \label{Eq:R_f}
    R_\text{f} = \frac{v_\text{ini}}{v_\text{ini} + v_{\text{orb,p}}},
\end{equation}
with a lower ratio $R_\text{f}$ predicting the formation of more elongated wind structures. The ratios for the circular binary models are provided in Table ~\ref{Ta:inputTableBinaries}, and are used in Sect.~\ref{ch:binSim}.

\section{Wind morphology in binary systems}
\label{ch:binSim}

Before analysing the impact of a third companion on the wind morphology, we investigate the outflow of the binaries to which our triple systems will be compared.

\begin{figure*}%[h!]
    \centering
    \includegraphics[width = 0.8 \textwidth]{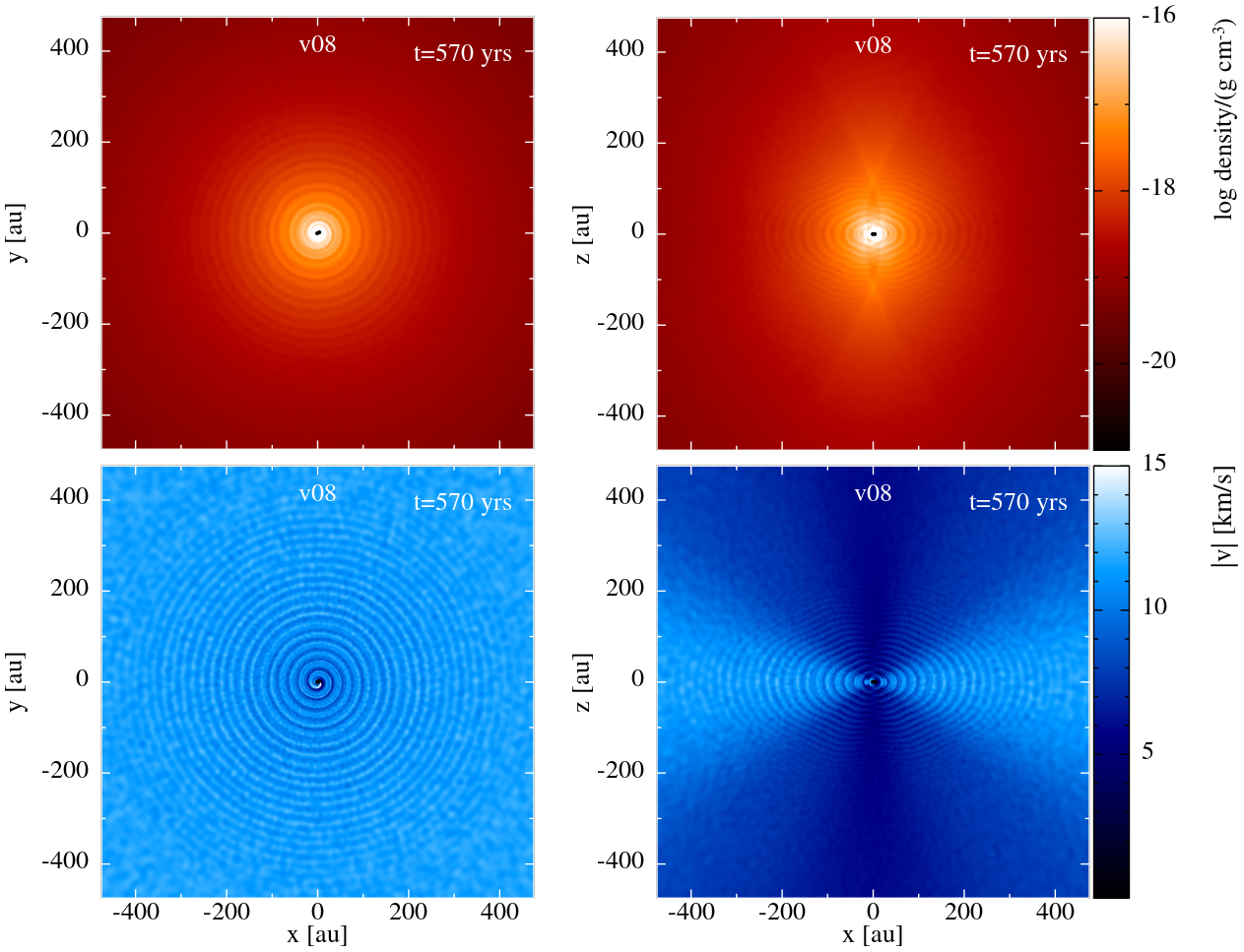}
    \caption{Density (top) and velocity (bottom) distribution in a slice through the orbital (left) and meridional (right) plane of model v08bin$_{\rm i}$ for $r\le 475 \, {\rm au}$.}
    \label{fig:clBin_v08_z475}
\end{figure*}

\begin{figure*}%[h!]
    \centering
    \includegraphics[width = 0.8 \textwidth]{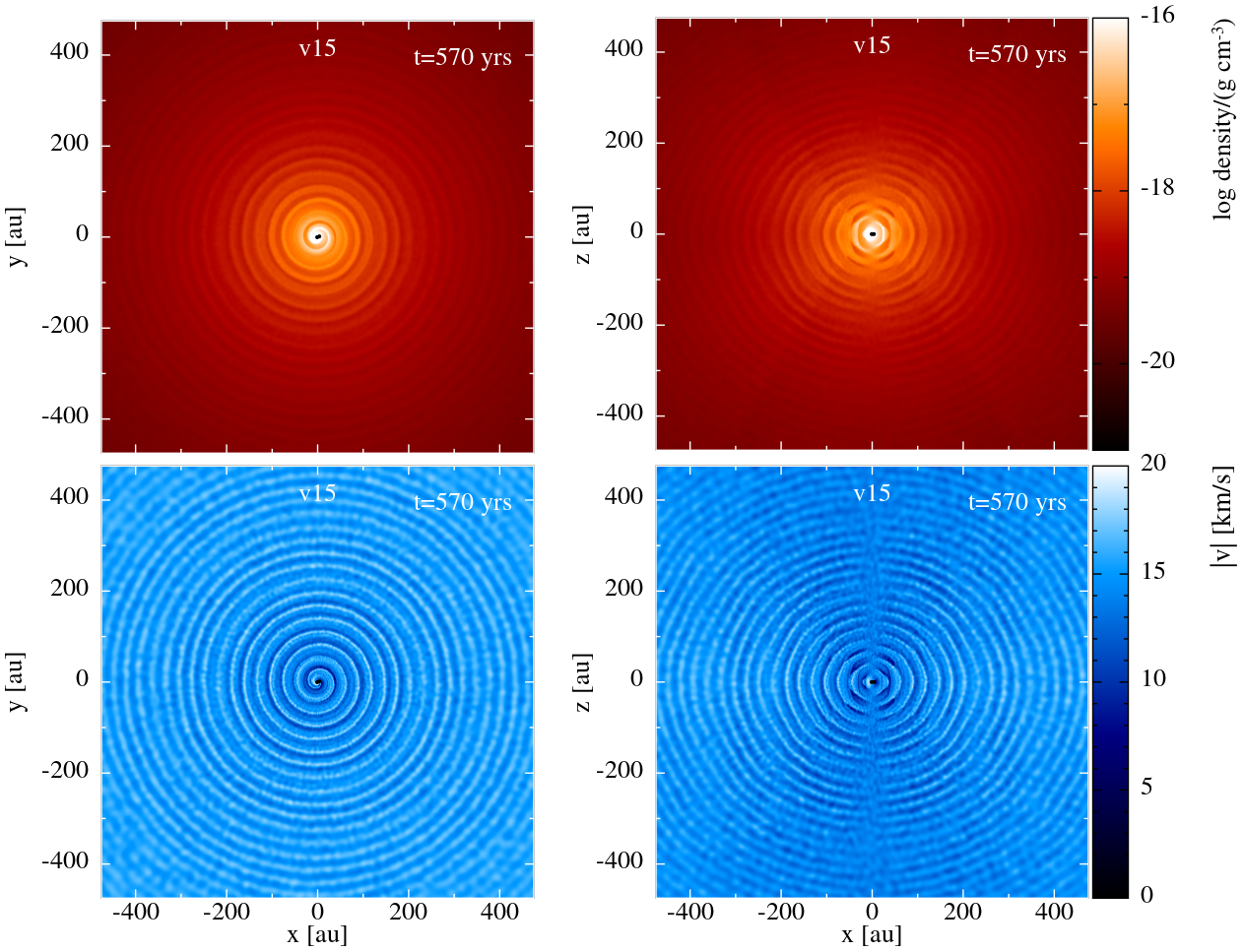}
    \caption{Same as Fig.~\ref{fig:clBin_v08_z475} for model v15bin$_{\rm i}$.}
    \label{fig:clBin_v15_z475}
\end{figure*}

First, we discuss the inner binary without outer companion, for initial wind velocities of $v_{\rm ini} = 8$ and $ 15 \, {\rm{km\,s^{-1}}}$ (models v08bin$_{\rm i}$ and v15bin$_{\rm i}$).
Figs.~\ref{fig:clBin_v08_z475} and ~\ref{fig:clBin_v15_z475} reveal the wind morphology of these systems, showing the density and velocity distribution in slices through the orbital (left panels) and meridional (right panels) planes. As is expected from the $\varepsilon$ values close to $1$ (Table~\ref{Ta:inputTableBinaries}), in both models an Archimedean spiral structure forms in the orbital plane, and concentric arcs appear in the edge-on plane. For model v15bin$_{\rm i}$ shown in Fig.~\ref{fig:clBin_v15_z475} these arcs resemble `banana'-like shapes with narrow ends that stretch towards the poles, and thicker 2-edged regions around the orbital plane ($z=0$). The thicker part of the banana-shapes emerges from the 2-edged gravity wake of the companion, which is shaped by the orbital motion of the AGB star and the direct gravitational attraction of the companion on the wind, while the narrow ends, reaching up to the poles, result solely from the orbital motion of the stars \citep{Kim2012B}. 
The transition between the thicker centres and narrower edges is visible as a straight line at an angle $\theta \sim 50\degree $ with respect to the orbital plane. 
The stronger the relative importance of the direct gravitational impact of the companion (compared to the impact of the orbital motion of the AGB), the larger the cross-section of the gravity wake will be, and the larger the angle of this transition line is expected. 
For model v08bin$_{\rm i}$, in which the lower velocity causes a stronger gravitational influence of the companion on the wind (larger $\varepsilon$ value, Table~\ref{Ta:inputTableBinaries}), the distinction between the thin edges and broader centres of the arc-like shapes can only be made for the innermost arcs at $r\lesssim 50 \, {\rm au} $. Further out, the cross-section of the gravity wake (so thicker part of the arcs) expands up to the poles. As a result, the lines resulting from the transition between the thin edges and broader centres, are not linear, but follow an increasing angle $\theta$ with respect to the orbital plane for increasing $r$. When the edges reach the poles ($\theta = \pi/2$), they interact with the structures from the opposite side (at $x=0$, $y\approx \pm 120 \, {\rm au}$).

Comparing the morphology in the meridional plane of both models, the arc structures in model v08bin$_{\rm i}$ are more elongated than in model v15bin$_{\rm i}$. 
In these models, the primary's orbital velocity, which causes a flattening of the wind morphology (see Sect.~\ref{ch:method:EpsFlR}), is $v_{\text{orb,p}} = 3.77 \, {\rm km \, s^{-1}}$, which is relatively low due to the low secondary mass $M_2 = 0.4 \, {\rm M_\odot}$.
The edge-on velocity distribution of model v08bin$_{\rm i}$ in Fig.~\ref{fig:clBin_v08_z475} (lower right panel) illustrates that there is a flattening due this orbital motion, as the velocity $|v|$ within the arc-structures decreases with increasing angle away from the orbital plane, with $|v| \approx 12 \, {\rm km \, s^{-1}}$ in the orbital plane, to $|v| \approx 6 \, {\rm km \, s^{-1}}$ close to the poles. 
The theoretical flattening of these models, from Eq.~\ref{Eq:R_f}, resulting solely from the impact of the orbital velocity on the velocity distribution in the wind, are $0.68$ and $0.80$, for models v08bin$_{\rm i}$ and v15bin$_{\rm i}$, respectively.
The actual flattening ratio in the simulations is estimated by measuring the height-to-width ratio of the edge-on arc-like structures in Figs.~\ref{fig:clBin_v08_z475} and ~\ref{fig:clBin_v15_z475}, and are $\sim 0.6$ and $\sim 0.8$ for v08bin$_{\rm i}$ and v15bin$_{\rm i}$.
This shows that the theoretical prediction of the flattening (which does not take into account the direct gravitational effect of the companion) is somewhat underestimating the flattening in model v08bin$_{\rm i}$, but is a good approximation for model v15bin$_{\rm i}$.

\begin{figure}%[h!]
    \centering
    \includegraphics[width = 0.4 \textwidth]{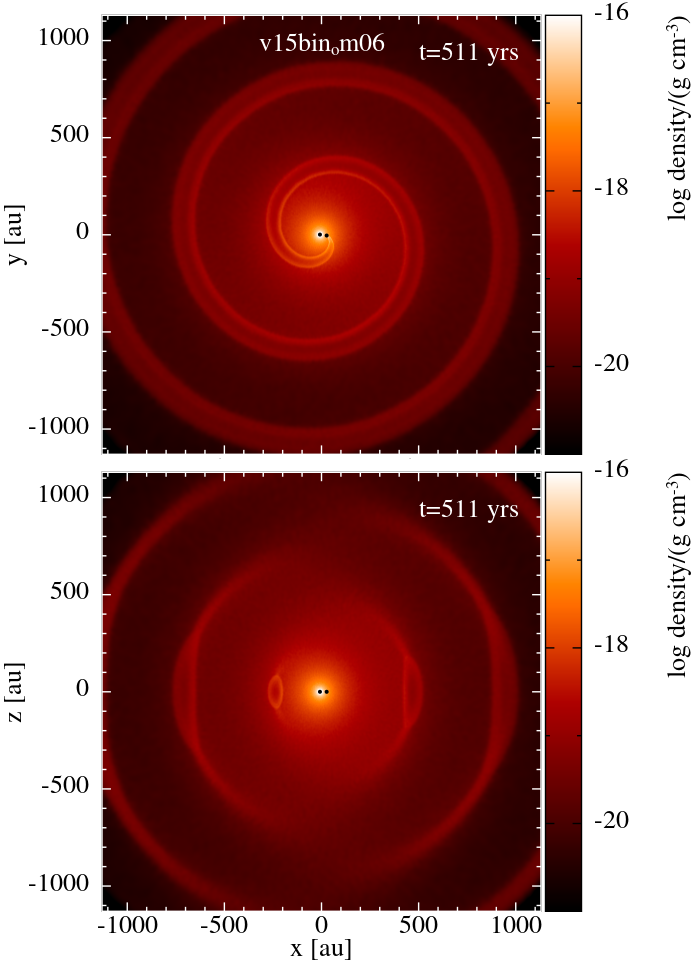}
    \caption{Density distribution in a slice in (top) and perpendicular to (bottom) the orbital plane of model v15bin$_{\rm o}$m06.}
    \label{fig:wideBin_z1130}
\end{figure}

Next, we present the outer binaries v15bin$_{\rm o}$m06 and v15bin$_{\rm o}$m06e40, which consist of a primary AGB star with $M_{\rm p} = M_{\rm 1} + M_{\rm 2}$ and outer companion $M_{\rm s} = M_{\rm 3} = 0.6$, an input wind velocity $v_{\rm ini} = 15 \, {\rm km \, s^{-1}}$, and eccentricities $0$ and $0.4$, respectively. Because of the large orbital separation, and thereby lower density $\rho$ at the location of the companion, its gravitational impact is expected to be smaller, which is in agreement with the smaller $\varepsilon < 1$ (Table~\ref{Ta:inputTableBinaries}).
Fig.~\ref{fig:wideBin_z1130} shows the morphology of model v15bin$_{\rm o}$m06. In the orbital plane, it consists of an extended face-on 2-edged Archimedian spiral density enhancement on top of a spherically symmetric wind. In the meridional plane, there are edge-on arcs with thin ends reaching towards the polar regions and thicker 2-edged structures in the centre that reach up to only $\theta \sim 25 \degree$ with respect to the orbital plane, due to the reduced gravitational pull of the companion.
The orbital velocity of the primary star, $v_{\text{orb,p}} = 1.87 \, {\rm km \, s^{-1}}$, is lower than in the inner binary models due to the large orbital separation. As a consequence, the morphology is less elongated, with a theoretical flattening ratio $R_{\rm f} = 0.89$ (Eq.~\ref{Eq:R_f}) and a measured ratio $R_{\rm f, sim} \approx 1$.
Additionally, due to the large orbital separation, and relatively high wind velocity, the spiral structure travels a significant distance during one orbital period, such that the distance between two consecutive arcs or spiral arms is larger compared to shorter period binary models. Unlike in the inner binary models, the structures do not interact and overlap in the spatial range of the simulation, and the quasi-spherical wind is still observable between the 3D structures. 
In the snapshot of Fig.~\ref{fig:wideBin_z1130}, the inner edge of the spiral wake crosses the $x$-axis at $x \approx 450, \, 900$ and  $1350 \, {\rm au}$, indicating that the structure travels $\sim 450 \, {\rm au}$ radially outward during one orbital period of $128.5 \, {\rm yrs}$, thereby moving with a velocity of approximately $16.6 \, {\rm km \, s^{-1}}$. 
In observational studies, the distance between two recurrent high-density structures can in this way be used to estimate the orbital period or wind velocity, under the condition that a good estimate of the other quantity is available.

\begin{figure}%[h!]
    \centering
    \includegraphics[width = 0.4 \textwidth]{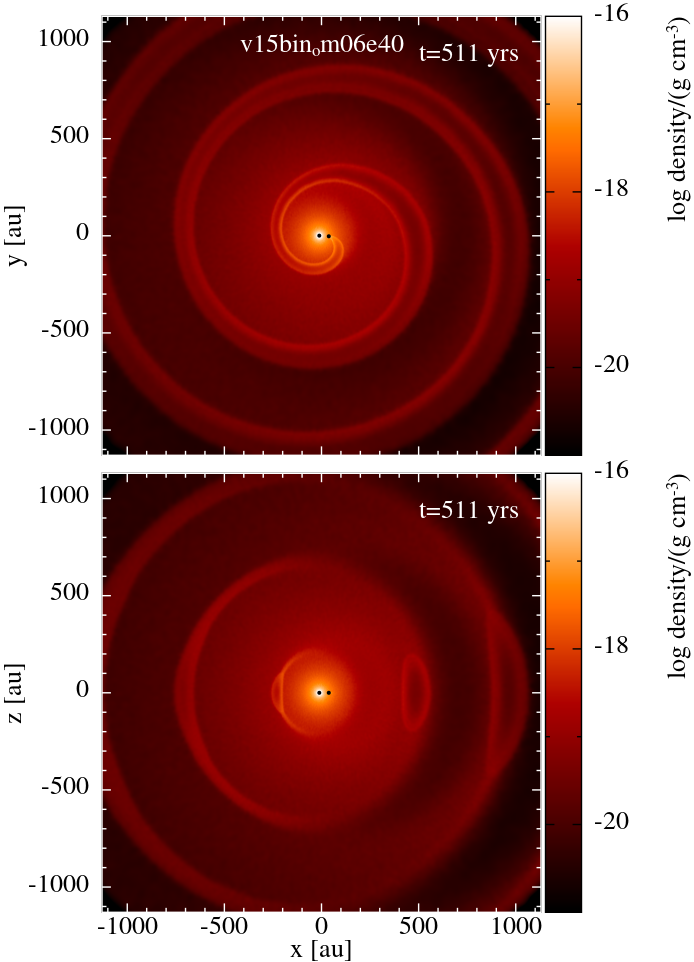}
    \caption{Density distribution in a slice in (top) and perpendicular to (bottom) the orbital plane of model v15bin$_{\rm o}$m06e40. The apastron passage of the companion is on the $y=0, x>0$ axis}
    \label{fig:wideBine40_z1130}
\end{figure}

Fig.~\ref{fig:wideBine40_z1130} exhibits the density distribution in the orbital and an edge-on plane of the eccentric system v15bin$_{\rm o}$m06e40. The inclusion of eccentricity results in a globally more asymmetric morphology. At the apastron side of the companion ($x>0$), the minimal orbital velocity results in a widening of the gravity wake, and at periastron side ($x<0$), the higher orbital velocity makes the wake thinner. As a consequence, the edge-on arcs at apastron side consist mainly of broad 2-edged structures centred around the orbital plane, and at periastron side of large, thin concentric arcs.


\section{Wind morphology in triple systems}
\label{ch:triples}

\begin{figure*}
    \centering
    \includegraphics[width = \textwidth]{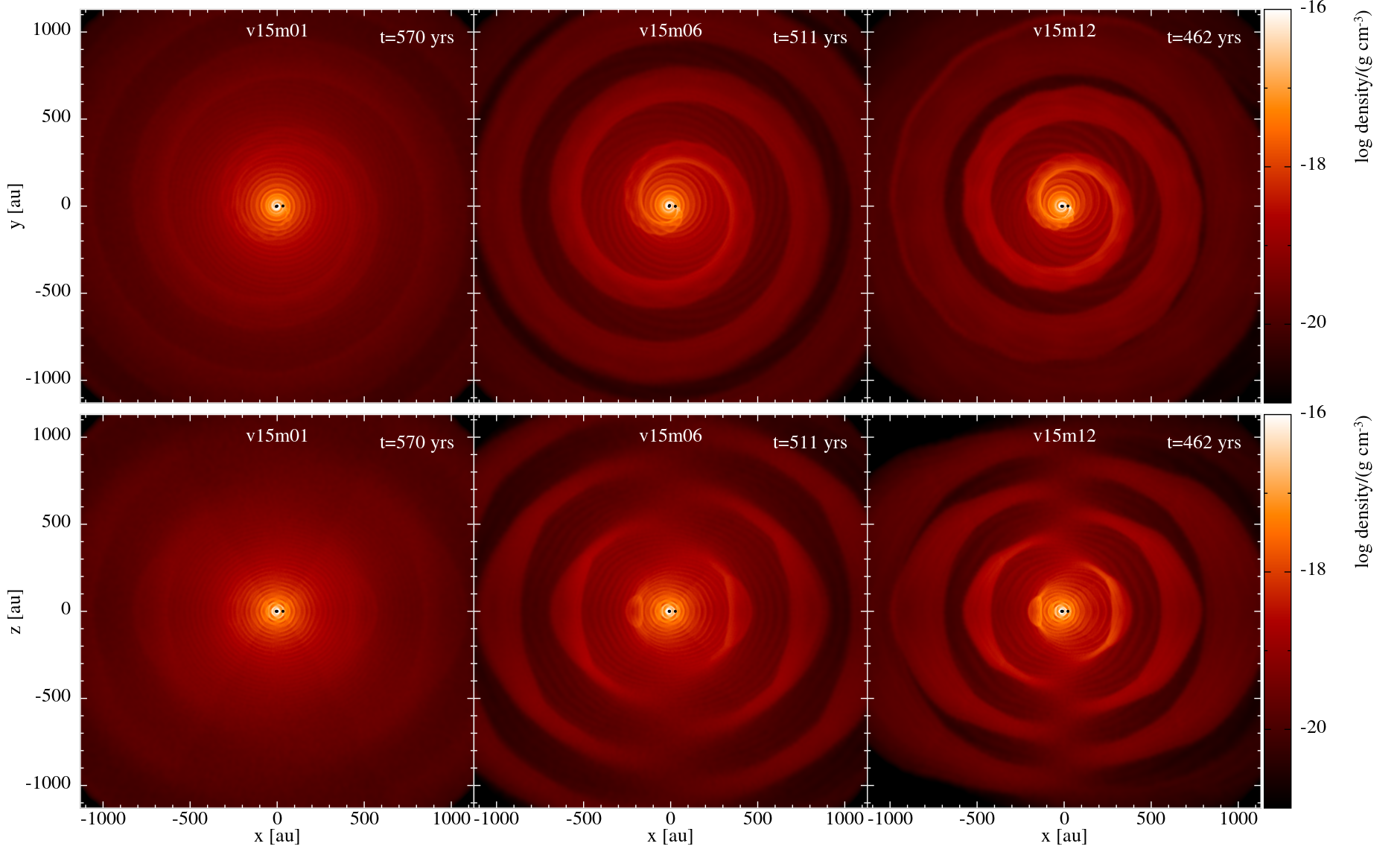}
    \caption{Density distribution in a slice in the orbital plane (top) and in a perpendicular meridional plane (bottom) for models v15m01 (left), v15m06 (middle), and v15m12 (right).}
    \label{fig:tr_v15_rho_z1130}
\end{figure*}

\begin{figure*}
    \centering
    \includegraphics[width = \textwidth]{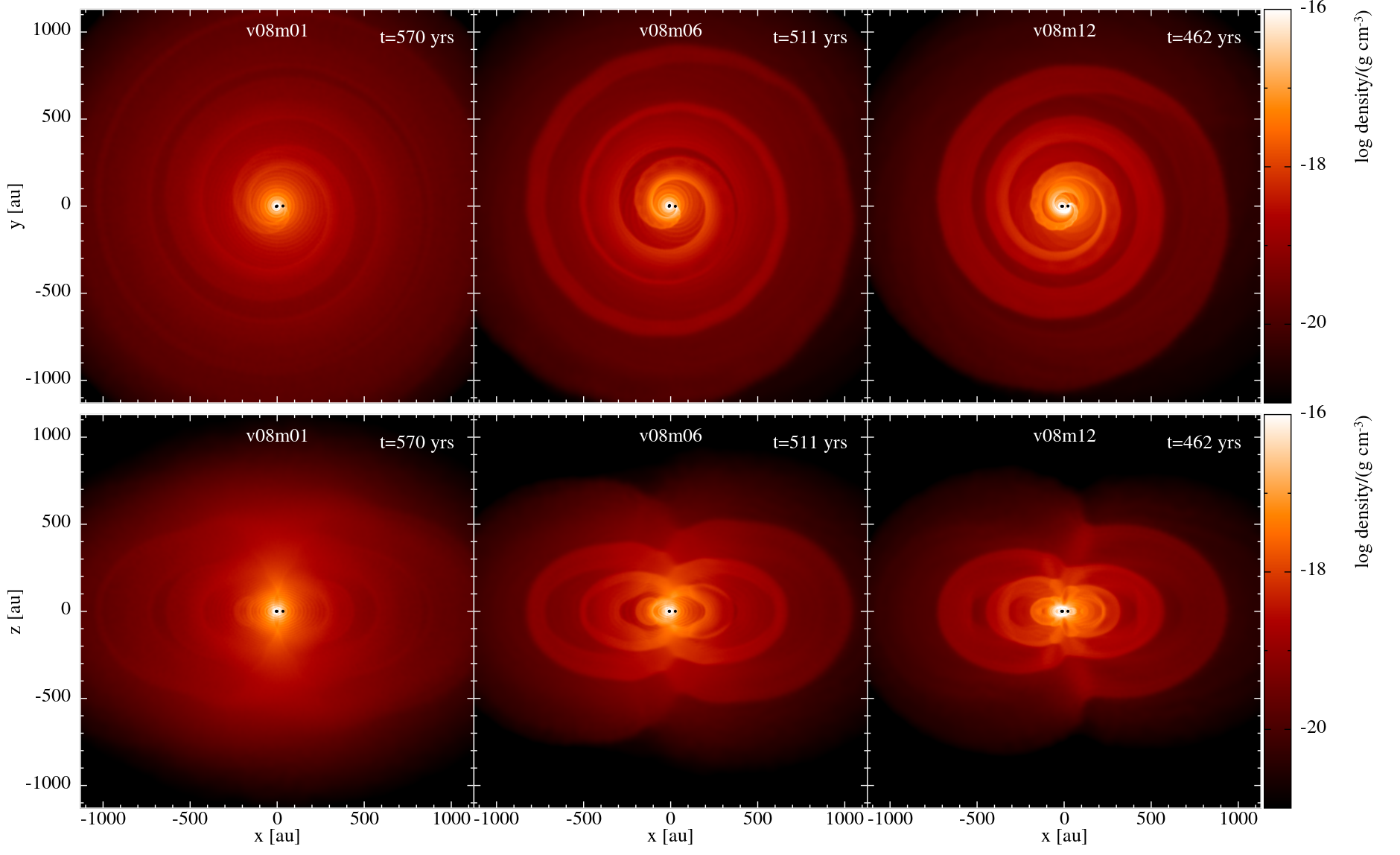}
    \caption{Same as Fig.~\ref{fig:tr_v15_rho_z1130} for models v08m01 (left), v08m06 (middle), and v08m12 (right).}
    \label{fig:tr_v08_rho_z1130}
\end{figure*}

\begin{figure*}[h!]
    \centering
    \includegraphics[width = 0.9\textwidth]{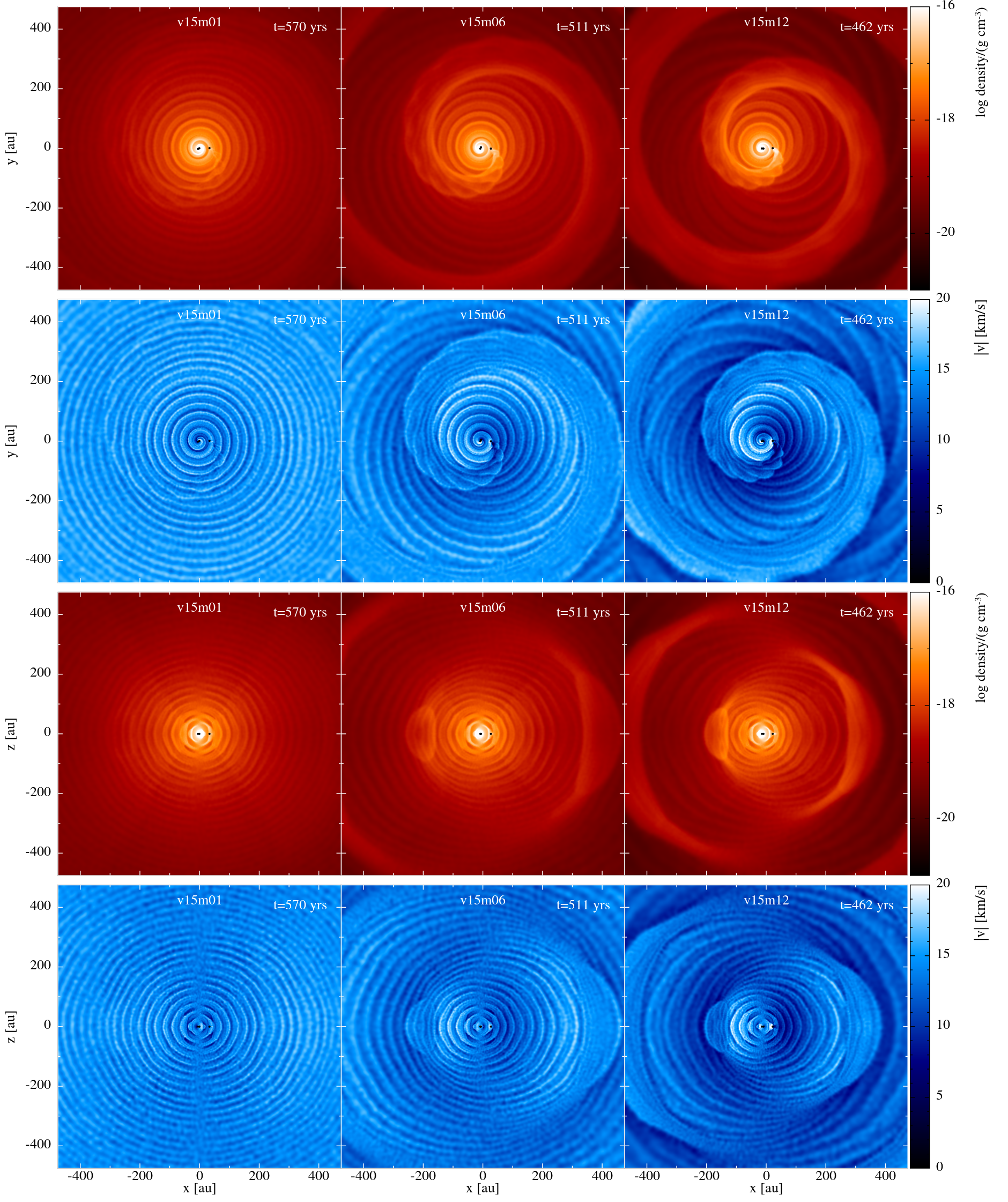}
    \caption{Density and velocity distribution in a slice in the orbital plane (top) and in a perpendicular meridional plane (bottom) for models v15m01 (left), v15m06 (middle), and v15m12 (right), for $r \le 475 \, {\rm au}$.}
    \label{fig:tr_v15_z475}
\end{figure*}

\begin{figure*}[h!]
    \centering
    \includegraphics[width = 0.9\textwidth]{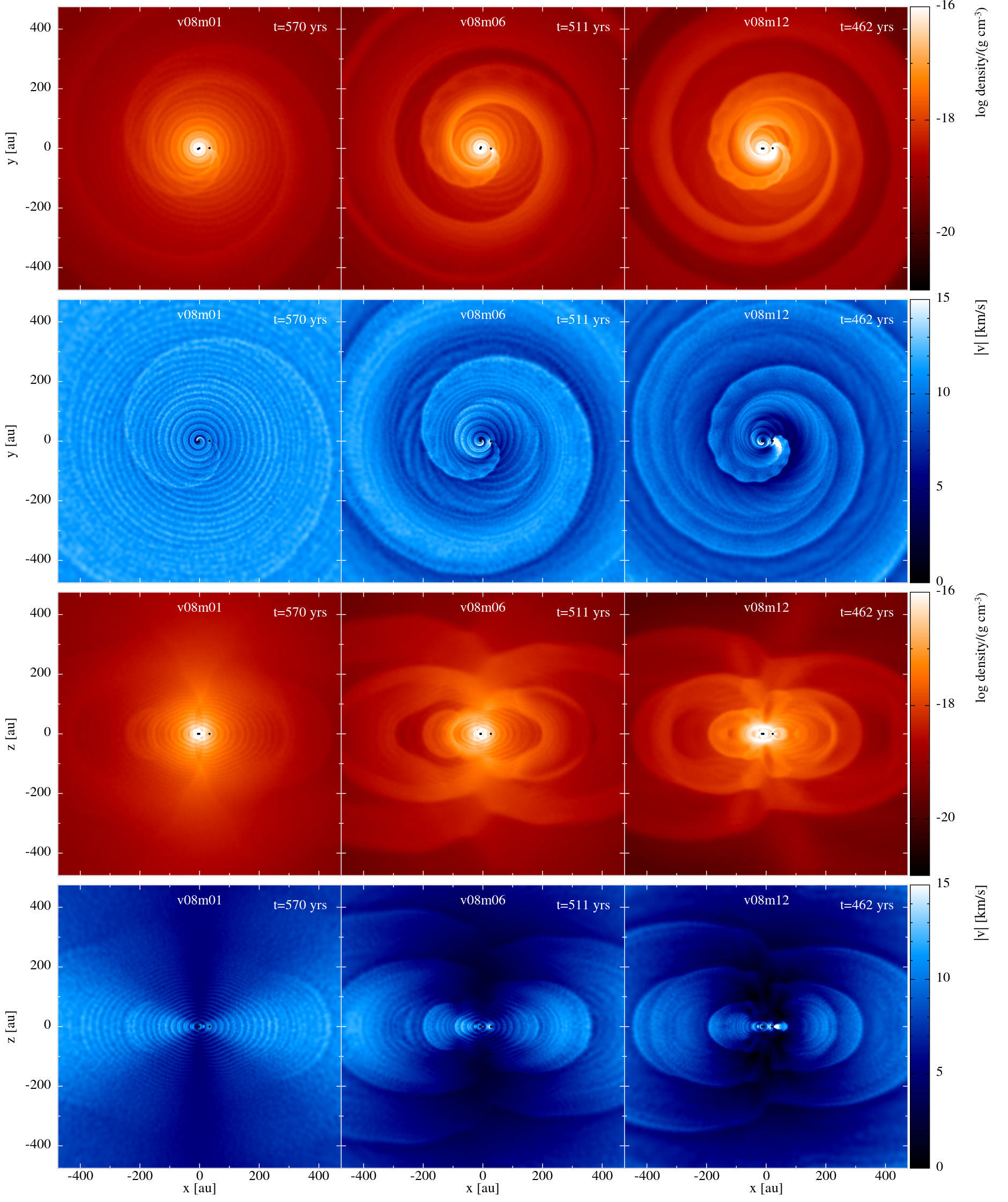}
    \caption{Same as Fig.~\ref{fig:tr_v15_z475} for models v08m01 (left), v08m06 (middle), and v08m12 (right).}
    \label{fig:tr_v08_z475}
\end{figure*}

We study the wind morphologies in hierarchical triple systems, using the set of models as defined in Fig.~\ref{Fig:hierarchTripleSetup} and Tables~\ref{Ta:setupTableTriples} and ~\ref{Ta:inputTableTriples}. We focus on the impact of changing the triple companion mass $M_3$, the wind velocity $v_{\rm ini}$, and the eccentricity $e$.
We expect from theory that a lower wind velocity, increased companion mass, and higher eccentricity will make the wind deviate stronger from spherical symmetry, which is reflected in the values of $\varepsilon$, given in Table~\ref{Ta:inputTableTriples}.
Figs.~\ref{fig:tr_v15_rho_z1130} and ~\ref{fig:tr_v08_rho_z1130} display the density distribution of the non-eccentric models with outer companion masses $M_3 = 0.1, 0.6, 1.2 \, {\rm M_\odot}$, in the orbital plane (top) and a meridional plane (bottom) for input velocity $v_{\rm ini} = 15 \, {\rm km \, s^{-1}}$ and $v_{\rm ini} = 8 \, {\rm km \, s^{-1}}$, respectively.
Additionally, a more zoomed-in view of the density and velocity distribution of these models are shown in Figs.~\ref{fig:tr_v15_z475} (v15) and ~\ref{fig:tr_v08_z475} (v08).
The eccentric models are represented in Figs.~\ref{fig:tv15m06ei40_z1130},~\ref{fig:tv15m06ei40_z475_2ts},\ref{fig:tv15m06eo40_z1130}, and \ref{fig:tv15m06eo40_4ts_z250}.

\subsection{Impact of hierarchical third body}
\label{ch:addingTripleEffects}

We study the impact of hierarchically adding a third body to the binary systems, described in Sect.~\ref{ch:binSim}, from two different points-of-view: (i) considering the close binary model v15bin$_{\rm i}$ where a third, outer companion is added, and (ii) considering the wide binary model v15bin$_{\rm o}$m06, where the primary is split up into a close binary. 

First, we consider the effect of adding a third, outer body. 
Comparing the density distribution in the triple models in Fig.~\ref{fig:tr_v15_rho_z1130} to the inner binary v15bin$_{\rm i}$ in Fig.~\ref{fig:clBin_v15_z475}, we see that the face-on spiral and edge-on circular arcs, created by the inner binary $M_1$--$M_2$, are still present and resolved, independent of the outer companion mass $M_3$. 
On top of this compact underlying structure, there is an additional gravity wake in the orbital plane originating from the companion $M_3$, that appears as large concentric arcs in the meridional plane. 

We find that the more massive $M_3$ and thus the larger $\varepsilon_{\rm o}$ (from left to right in Fig.~\ref{fig:tr_v15_rho_z1130}), (i) the larger the density enhancement of its gravity wake and arcs with respect to the underlying wind and compact spiral pattern of $M_1$--$M_2$, and (ii) the more radially compressed its spiral and arcs are, because its orbital period is shorter, and the radial velocity within its gravity wake is lower. 
Although $\varepsilon_{\rm o}$ is very small for all outer binary systems, the outer companion $M_3$ is strongly altering the global 3D morphology.

Fig.~\ref{fig:tr_v15_z475} reveals in more detail the onset of the gravity wake spiral originating from $M_3$, through a more zoomed-in view of the density and velocity distribution in the orbital and edge-on plane of the triple v15 models. Because the outer companion on its orbit continuously encounters outward moving density- and velocity-enhanced spiral shocks from the inner binary, the outer edge of the $M_3$ gravity wake has a waved pattern. The orbital plane velocity and density distribution of model v15m06 (Fig.~\ref{fig:tr_v15_z475}, central column) illustrate more clearly how shocks propagate through this gravity wake, resulting in the waved outer edge. 
This waved pattern and the shocks within the gravity wake become less prominent further out in the wind. 
This is less noticeable in the edge-on view, but still visible, mainly in the velocity distribution of model v15m12 (lower right plot in Fig.~\ref{fig:tr_v15_z475}).

The companion $M_3$ also affects the shape and properties of the underlying spiral created by $M_1$--$M_2$. Because it induces an orbital motion of $M_1$--$M_2$ around the system centre-of-mass ${\rm{CoM}}_{M1+M2+M3}$, the orbital velocity of the AGB star and inner companion have an additional component, which affects the velocity and morphology of the $M_1$--$M_2$ spiral. The effect is stronger for a more massive outer companion $M_3$, and is thereby best visible in model v15m12. In the orbital plane and edge-on velocity plots of this model in Fig.~\ref{fig:tr_v15_z475} (right column), material in front of the inner binary ($x \lesssim  0, y \lesssim 0$) has significantly larger velocities than the material directly behind it and further back on its orbit ($y> 0$ and $x < 0$). 
Due to this orbital motion, the $M_1$--$M_2$ spiral appears as a snail-shell pattern.

Next, we discuss the effect of splitting the primary star of the wide binary v15bin$_{\rm o}$m06 into a close binary, resulting in the triple model v15m06. 
Comparing the $M_3$ gravity wake, in the middle panels in Fig.~\ref{fig:tr_v15_rho_z1130}, with the gravity wake in the wide binary in Fig.~\ref{fig:wideBin_z1130}, we see that the interaction of the companion $M_3$ with the inner spiral of $M_1$--$M_2$ has several effects. It not only (i) generates the wave-pattern in the outer edge of the $M_3$ gravity wake, but it also (ii) produces a broadening of this wake, and furthermore, (iii) the presence of the inner binary and its spiral pattern reduce the radial velocity of the $M_3$ gravity wake. 
This can be seen by comparing the radial distances travelled by the recurrent arcs and spiral arms in both models, and from the velocity profiles in Fig.~\ref{fig:tr_v15_z475} (model v15m06, second row, middle column) and Fig.~\ref{fig:wideBin_vel_z475}. For example, the inner and outer edge, respectively, of the third spiral and arc on the $x<0$ side have travelled about $850$ and $1100 \, {\rm au}$ in model v15m06, and about $ 1100$ and $ 1250 \, {\rm au}$ in the binary model v15bin$_{\rm o}$m06.

\subsection{Impact of different wind velocity}
Decreasing the wind velocity in triple systems increases the gravitational impact of both companions (see also lower $\varepsilon_{\rm i}, \varepsilon_{\rm o}$ in Table~\ref{Ta:inputTableTriples}). The effect on the global morphology can be seen by comparing the density distribution of the three models with different mass $M_3$ in Fig.~\ref{fig:tr_v15_rho_z1130} (v15) and Fig.~\ref{fig:tr_v08_rho_z1130} (v08), and at smaller scales in the density and velocity distribution in Fig.~\ref{fig:tr_v15_z475} (v15) and Fig~\ref{fig:tr_v08_z475} (v08).
In the edge-on view, there is a stronger flattening in case of lower wind velocity. The cross-section of the spirals produced by $M_3$ can no longer be described as concentric banana-shaped arcs, but resemble a bit more bicentric elongated structures that extend up to the polar axes. Similar to the binary model v08bin$_{\rm i}$ (Fig.~\ref{fig:clBin_v08_z475}), the edge-on velocity distribution in Fig.~\ref{fig:tr_v08_z475} reveals the characteristic decrease in velocity with increasing angle with respect to the orbital plane, that appears in flattened morphologies.
The flattening ratio of the inner binary model v08bin$_{\rm i}$ was measured as $R_{\rm f,sim} \approx 0.6$.
In the triple v08 models, these ratios are determined by the edge-on arcs originating from the triple companion $M_3$, as long as the density contrast of these arcs with the underlying wind is large enough to be visible and measurable. This measured $R_{\rm f}$ is approximately $0.5$ for all masses of $M_3$.
This indicates that compared to the binary models, the triple companion slightly increases the elongation of the morphology, but varying its mass does not significantly affect the degree of flattening, at least in models where the density contrast is still visible.

As in the v15 models, the orbital plane morphology of the v08 models is again characterised by a large gravity wake spiral on top of the underlying compact binary spiral pattern. The density distribution in Fig.~\ref{fig:tr_v08_z475} reveals that the gravity wake of $M_3$ in models v08m01 and v08m06 is a 2-edged spiral, with a wave-pattern in the outer edge.
The structure in model v08m12 is different, with a bow shock originating in front of $M_3$, instead of a 2-edged gravity wake behind it. 
We first focus on comparing models v08m01 and v08m06 in Fig.~\ref{fig:tr_v08_z475} with their corresponding v15 triple models in Fig.~\ref{fig:tr_v15_z475}, before going into more detail about model v08m12.
Firstly, the waves in the outer edge of the $M_3$ gravity wake are less pronounced in the v08 models because the interaction of the inner $M_1$--$M_2$ spiral shocks with $M_3$ and its gravity wake is less strong. This is, firstly, because the inner binary spiral shocks have lower velocities when they collide with the $M_3$ wake, as can be seen by comparing the orbital plane velocity distribution in Fig.~\ref{fig:tr_v08_z475} and ~\ref{fig:tr_v15_z475}. We note that the colorbars in these figures have different limits. Secondly, the density contrast encountered by $M_3$ as it passes a $M_1$--$M_2$ spiral shock is much smaller than in case of the higher wind velocity models, because in the v08 models, the inner edge of the $M_1$--$M_2$ spiral has already caught up with the outer edge at the orbital radius of the companion, as can be seen in the zoomed in density distribution of models v08m06 and v15m06 in Fig.~\ref{fig:v08_v15_m06_z100}.
Thirdly, when the wind has a lower velocity, the $M_3$ gravity wake is also wider (the distance between the inner and outer edge is larger), and is wrapped more closely around the system CoM due to the lower radial velocity of the material within the wake. As a result, the outer edge of the $M_3$ gravity wake catches up and merges with the inner edge that originated one orbital period earlier, after respectively $\sim 2$ and $1$ orbital periods in models v08m01 and v08m06. Beyond this interaction radius, the spiral pattern of the $M_1$--$M_2$ binary is no longer visible. From this, we conclude that the lower the wind velocity, and the higher the mass $M_3$, the lower the radius up to where the $M_1$--$M_2$ spiral pattern can be resolved. 

Out of all systems, the highest $\varepsilon_{\rm i}$ and $\varepsilon_{\rm o}$ values are found in model v08m12, indicating that the impact of the companions on the wind morphology is strongest.
Due to the cumulative effect of the relatively large companion mass $M_3$ and low wind velocity, the companion $M_3$ is able to gravitationally attract enough material around it to form an accretion disk and a bow shock. Accretion disks are expected to form around the AGB companion(s), but to resolve them numerically requires very small accretion radii and a very high number of SPH particles \citep{Malfait2024}. These conditions cannot be reached in our triples setups because of the large radial extent of the systems.
We find an accumulation of wind material around and behind the companion $M_3$ of model v08m12, such that after $\sim 300 \, {\rm yrs}$ of simulation time the wind morphology starts to change. This effect is visible through the high-density region behind the outer companion in the orbital plane density distribution in the upper right panel of Fig.~\ref{fig:tr_v08_z475}.
We performed a test simulation of this model with $\mu = 1.26$, which is more suited for atomic gas, and find no accumulation of material. Additionally, this change of $\mu$ increases the wind velocity throughout the entire wind, which strongly affects the entire morphology, as was concluded as well by \citep{Malfait2024}.
This indicates that a more robust treatment of the equation of state that takes into account the variations of $\mu$ and $\gamma$ is needed.

\subsection{Impact of eccentricity}

\begin{figure}[h!]
    \centering
    \includegraphics[width = 0.49\textwidth]{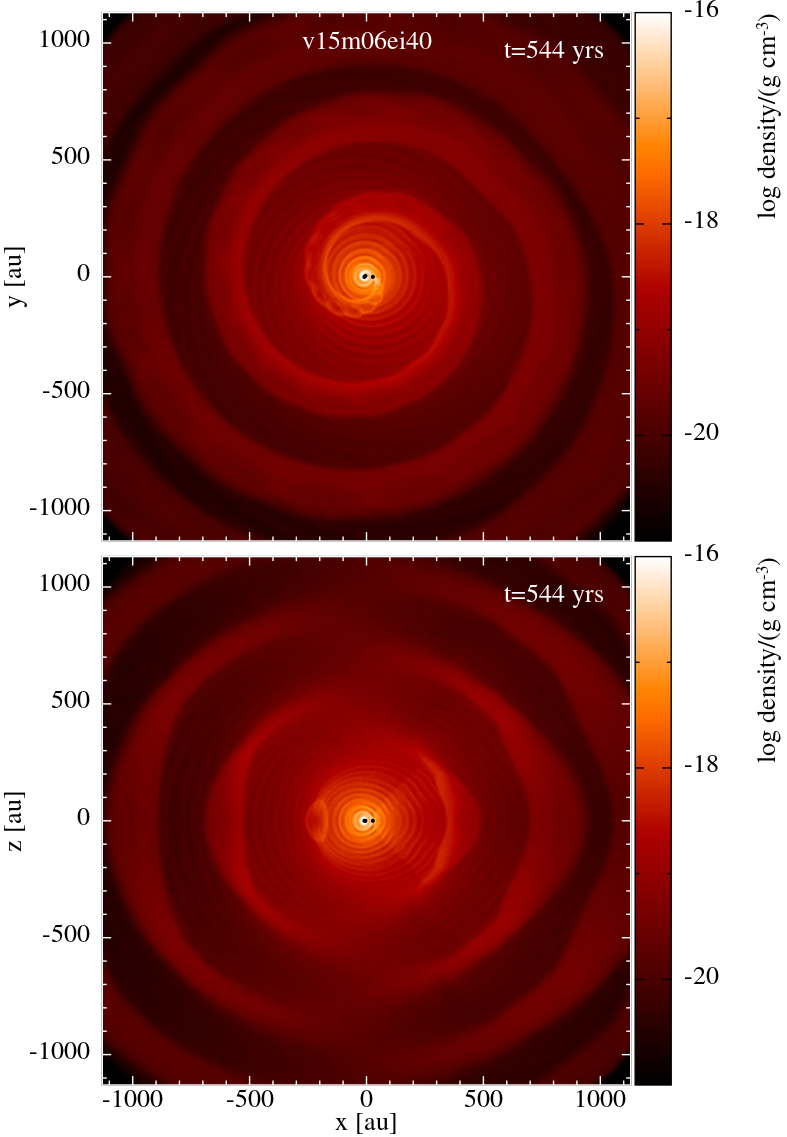}
    \caption{Density distribution in a slice in the orbital plane (top) and in a perpendicular meridional plane (bottom) for model v15m06e$_{\rm i}$40, with apastron passage of the inner companion at $x>0, y=0, z=0$.}
    \label{fig:tv15m06ei40_z1130}
\end{figure}
\begin{figure*}[h!]
    \centering
    \includegraphics[width = 0.7\textwidth]{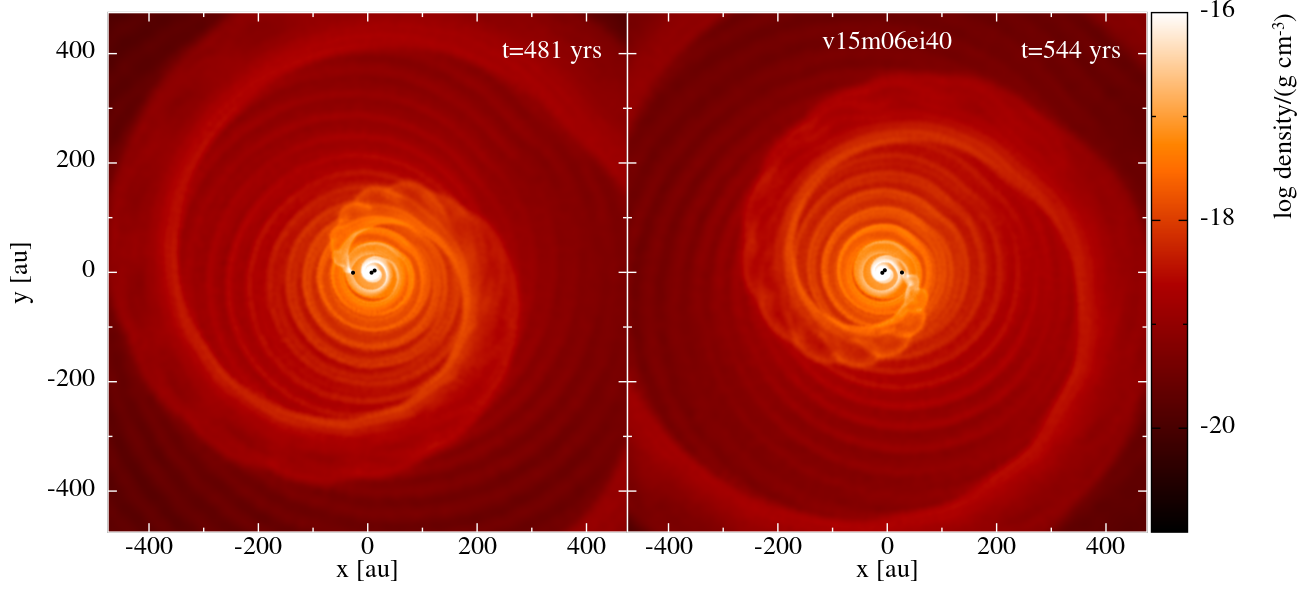}
    \caption{Density distribution in a slice in the orbital plane at $2$ different orbital phases for model v15m06e$_{\rm i}$40. The apastron passage of the inner companion star $M_2$ is on the $x>0, y=0, z=0$ axis. }
    \label{fig:tv15m06ei40_z475_2ts}
\end{figure*}

To study the characteristics of eccentricity in triple system morphologies, an eccentricity of $0.4$ is introduced to the inner and outer orbit in models v15m06e$_{\rm i}$40 and v15m06e$_{\rm o}$40, respectively.
For the model with an inner eccentric orbit, the global density distribution (Fig.~\ref{fig:tv15m06ei40_z1130}) is very similar to the one of the non-eccentric model v15m06 of Fig.~\ref{fig:tr_v15_rho_z1130} (middle row). The largest difference is that at apastron side ($x>0$), the spiral wake of $M_3$ is slightly distorted. This distortion is revealed as an asymmetry in the meridional plane slice (through $y=0$, Fig.~\ref{fig:tv15m06ei40_z1130}). The edge-on arcs have a different shape at apastron side ($x>0$), with a thicker central bulge around the orbital plane, than at periastron side ($x<0$).
The effect of the eccentric inner orbit is more prominent on smaller scales, where it induces an asymmetry in the compact spiral of $M_1$--$M_2$ that is visible between the windings of the outer $M_3$ spiral wake (see Fig.~\ref{fig:tv15m06ei40_z475_2ts}). 
This asymmetric compact $M_1$--$M_2$ spiral deviates from the one in the non-eccentric model v15m06 (Fig.~\ref{fig:tr_v15_z475}), especially at the upper right side of the orbital plane ($x>0, y>0$), in a very similar way as model v20e50 in \cite{Malfait2021,Malfait2024}, where the origin of this type of asymmetry is explained in detail.
We conclude that the eccentricity of the inner orbit has a limited effect on the large-scale wind structures, but is revealed on a smaller-scale, indicating that this type of eccentricity can probably only be detected observationally if the inner wind structures are resolved well. 

\begin{figure}[h!]
    \centering
    \includegraphics[width = 0.49 \textwidth]{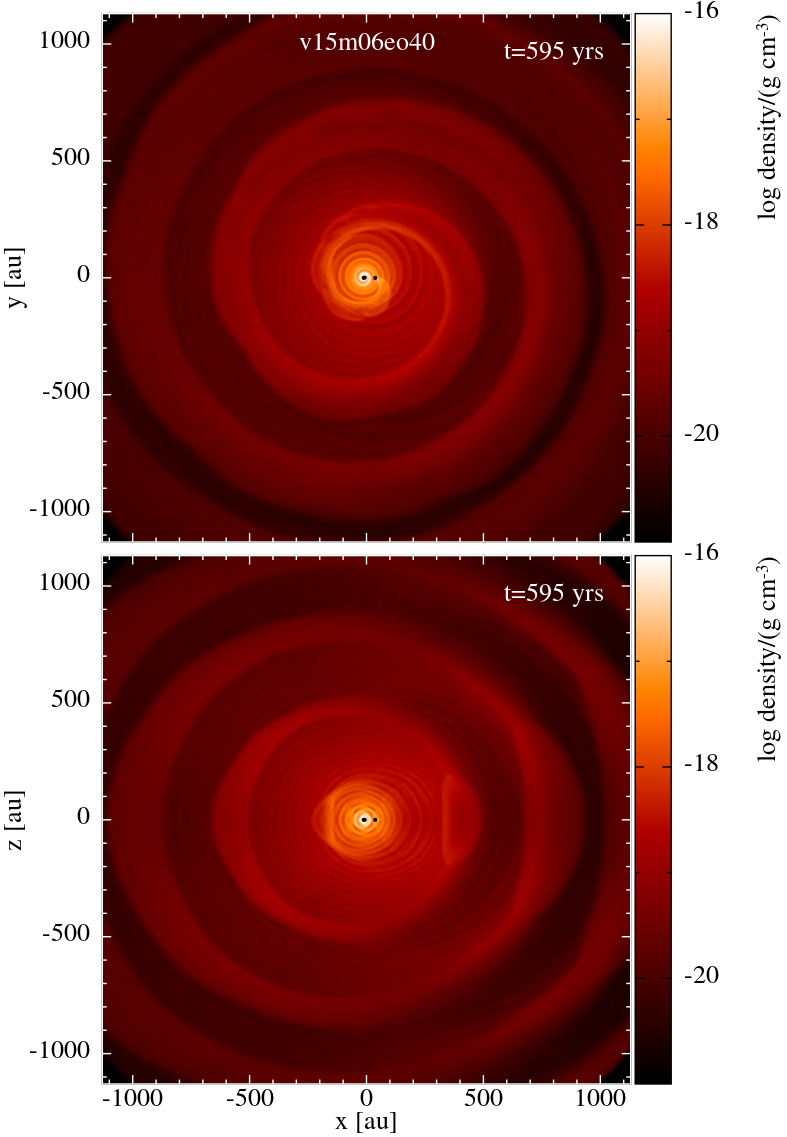}
    \caption{ Density distribution in a slice in the orbital plane (top) and in a perpendicular meridional plane (bottom) for model v15m06e$_{\rm o}$40, with apastron passage of the outer companion at $x>0, y=0, z=0$.
    }
    \label{fig:tv15m06eo40_z1130}
\end{figure}

\begin{figure*}[h!]
    \centering
    \includegraphics[width = \textwidth]{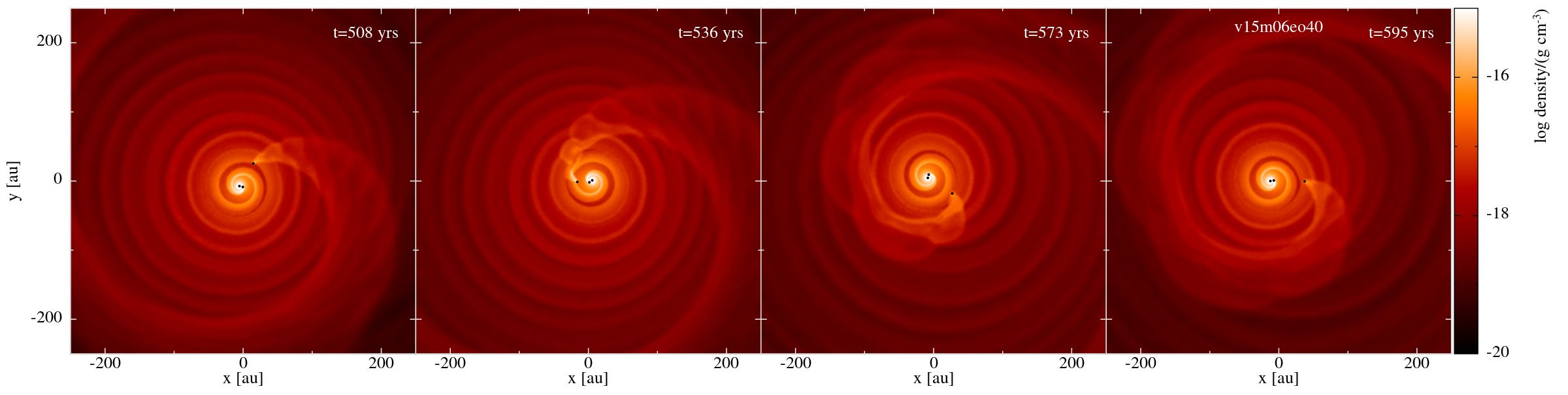}
    \caption{ Density distribution in a slice in the orbital plane at $4$ different orbital phases within the outer orbital period of model v15m06e$_{\rm o}$40. The companion star $M_3$ is at periastron passage in the second image ($x<0$,$y=0$) and at apastron passage in the fourth image ($x>0, y=0, z=0$). }
    \label{fig:tv15m06eo40_4ts_z250}
\end{figure*}

The impact of the eccentricity on the large-scale wind morphology is more pronounced if it is induced in the outer orbit, as can be seen from the density distribution in the orbital and edge-on planes in Fig.~\ref{fig:tv15m06eo40_z1130} of model v15m06e$_{\rm o}$40. 
In the orbital plane, the outer spiral wake is thicker at apastron side ($x>0$) and more narrow at periastron side ($x<0$).
In the meridional plane, the periastron side contains extended arcs reaching up to the polar regions, whereas at apastron side, the structures resemble more banana-shapes with thicker 2-edged bulges more confined around the orbital plane.
The asymmetries induced by eccentricity are less obvious than in the eccentric close binary systems studied by \cite{Malfait2021} and \cite{Malfait2024}.
To investigate whether this is purely due to the larger extent of the orbit, or due to the added complexities induced by the triple nature of the system, we compare this model (v15m06e$_{\rm o}$40) to the eccentric binary model v15bin$_{\rm o}$m06e40 in Fig.~\ref{fig:wideBine40_z1130}.
We conclude that the previously mentioned asymmetries are more pronounced in the binary model, and are thereby indeed affected by the triple nature of the system.
When we split up the primary into a close binary, similar effects result as in the comparison between the circular triple model v15m06 and the wide binary v15bin$_{\rm o}$m06 in Sect.~\ref{ch:addingTripleEffects}.
Due to the triple nature of the system, there is (i) a waved pattern on the outer edge of the gravity wake of $M_3$, this time particularly around periastron side, (ii) a broadening of the gravity wake, resulting in thicker arcs in the edge-on view, and (iii) a reduced radial velocity of the wake, making it travelling a shorter radial distance in the same number of orbital periods.
The aforementioned effects render it more challenging to discern the eccentric nature of a triple system when examining its wind morphology, in comparison to a binary system.

To study in more detail how the eccentricity-induced asymmetries arise in model v15m06e$_{\rm i}$40, Fig.~\ref{fig:tv15m06eo40_4ts_z250} shows the onset of the gravity wake of $M_3$ at four different orbital phases. Understanding the wind-companion interaction throughout the orbit is very challenging due to the many interferences and interacting components.
Ignoring the inner binary spiral shock of $M_1$--$M_2$, the changing orbital velocity of the outer companion $M_3$ (both in absolute value and direction) strongly affects the opening angle and width of the $M_3$ gravity wake, resulting in the asymmetries revealed as well in the outer binary model v15bin$_{\rm o}$m06e40 in Fig.~\ref{fig:wideBine40_z1130}. 
Further, the interaction of the companion $M_3$ with the spiral shocks created by the inner binary $M_1$--$M_2$ strongly determines the shape of the resulting gravity wake of $M_3$. The time in-between two encounters of $M_3$ with a spiral shock becomes phase-dependent because of the varying orbital velocity and separation along the orbit, and the varying distance between two $M_1$--$M_2$ spiral shocks.
Further, the impact of an encounter of $M_3$ with a passing spiral shock depends on the density contrast of the shock with the wind. This contrast is larger when the outer edge of the $M_1$--$M_2$ compact binary spiral has not yet caught up and merged with the previous inner edge. In this model, because of the varying orbital separation the $M_3$ companion sometimes resides within and sometimes beyond the region where these edges merge.
Finally, the shape of the $M_1$--$M_2$ binary spiral is affected by the phase-dependent orbital velocity of the CoM$_{M_1+M_2}$, and by the direct gravitational attraction of wind material towards companion $M_3$.

Taking into account these complexities, we attempt to understand the different shapes of the $M_3$ gravity wake at different orbital phases in Fig.~\ref{fig:tv15m06eo40_4ts_z250} as follows.
After apastron passage (first plot, $t = 508 \, {\rm yrs}$), the companion $M_3$ is accelerating, which results in a decrease of the width of its gravity wake. 
Because $M_3$ has a radially inward orbital velocity component, it encounters successive outward moving $M_1$--$M_2$ spiral shocks at a faster rate. 
Additionally, between apastron passage ($x>0, y=0$) and this orbital phase, the orbital radius of $M_3$ is relatively large, such that it resides in a region where the inner and outer edge of the $M_1$--$M_2$ spiral already merged, and the density contrast of the passing spiral shocks with the wind is relatively low.
As a consequence of these two effects, rather small waves result on the outer edge of the gravity wake, that fade out at larger radii (Fig.~\ref{fig:tv15m06eo40_z1130}).
In contrast, around periastron passage (second plot, $t = 536 \, {\rm yrs}$), $M_3$ resides in a region where the edges of the $M_1$--$M_2$ spiral have not yet merged, such that the density contrast of the passing spiral shocks and the wind is larger, which results in more impact-full collisions. 
After periastron, the radially outward directed orbital velocity reduces the rate at which the companion encounters spiral shocks. Further, it encounters strong spiral shocks, because it takes some time before it enters the region where the $M_1$--$M_2$ spiral edges have merged.  This results in the large waves in the outer edge of the gravity wake that can be seen in the third plot of Fig.~\ref{fig:tv15m06eo40_4ts_z250} (e.g.\,the wave at $-120 \, {\rm au} \lesssim x \lesssim 0  \, {\rm au}$ and $-100 \, {\rm au} \lesssim y\lesssim - 50 \, {\rm au}$ at $t=575$~yr).
As the companion approaches apastron, its orbital velocity direction becomes perpendicular to the radial velocity of the spiral shocks, which increases the rate at which spiral shocks pass by. Additionally, the encounters are less violent due to the larger orbital separation, such that the outer gravity wake edge becomes smoother again.
This explains why the gravity wake in the global orbital plane morphology in Fig.~\ref{fig:tv15m06eo40_z1130} is more wavy at periastron side ($x<0$), and smoother at apastron side ($x>0$).

\section{Comparison to R Aql}
\label{ch:RAqlSection}
Observations of the circumstellar envelope of the AGB star R Aql reveal a very high degree of complexity, indicating that one or multiple, still undetected, companions are probably shaping the wind \citep{Decin2020}. This inspired us to examine the possibility that a triple stellar system is embedded and hidden within this outflow, and to explore which type of wind structures and global morphologies are expected to form around such triple systems. We therefore modelled the set of systems, described and analysed in Sect.~\ref{ch:SimulationsMethod} and Sect.~\ref{ch:triples}, in such a way that the primary AGB star has similar properties as R Aql.

The exact nature of any companions of R Aql and the orbital properties are still unknown. Therefore, the parameter space that could be explored to model R Aql is extremely large and degenerate. 
However, in this section we perform a preliminary comparison of the observations of R Aql to synthetic observations of one or our triple models that lies within this degenerate parameter space.
A thorough study of the observational data, including different molecular lines, which all provide different insights on the possible nature of its central system and on the complex envelope at different spatial scales within the wind, is beyond the scope of this work {and is needed before a more detailed comparison with a simulation can be made}. To describe the observed morphology, we solely focus on the $^{12}$CO $J=2$\textrightarrow$1$ rotational line emission, which is an optimal tracer of the density in the entire envelope, due to its large fractional abundance with respect to molecular hydrogen, and because it is predominantly collisionally excited \citep{DeBeck2010}. 
We first shortly describe this observational data, and then examine what we can learn from a comparison to a synthetic observation of one or our models.

\subsection{ALMA observation}
\label{ch:ObservRAql}
R Aql is an O-rich AGB star that is located at a distance of $230\, {\rm pc}$ from Earth. 
Its mass is estimated to $1.6 \, M_\odot$ and its mass-loss rate is $\sim \dot{M} = 1.1 \times 10^{-6} \, {\rm{M_\odot\,yr^{-1}}}$ \citep{Decin2020}. 
R Aql was observed as part of the \textsc{ATOMIUM} ALMA large programme, with three different ALMA antenna array configurations: compact, mid and extended (with baseline ranges of $15-500, 15-1398,$ and $111-13894 \, {\rm m}$ respectively) \citep{Decin2020,Gottlieb2022}. To optimally reveal both small- and larger-scale structures, a combination is made of the data from these three configurations (with a $uv$ taper of FWHM $0.04 "$, which is equivalent to a baseline of $\sim 9 \, {\rm km \, s^{-1}}$ at $1.3 \, {\rm mm}$ wavelength, with restoring beams (major axis [mas], minor axis [mas], angle [degrees]), of ($40.7, 36.0, 22.38$) and ($58.3, 49.3, 25.12$) for the continuum and CO cube, respectively. 
We refer to \cite{Gottlieb2022} for more details on the observations and data processing. 

\begin{figure}
	\includegraphics[width = 0.5 \textwidth]{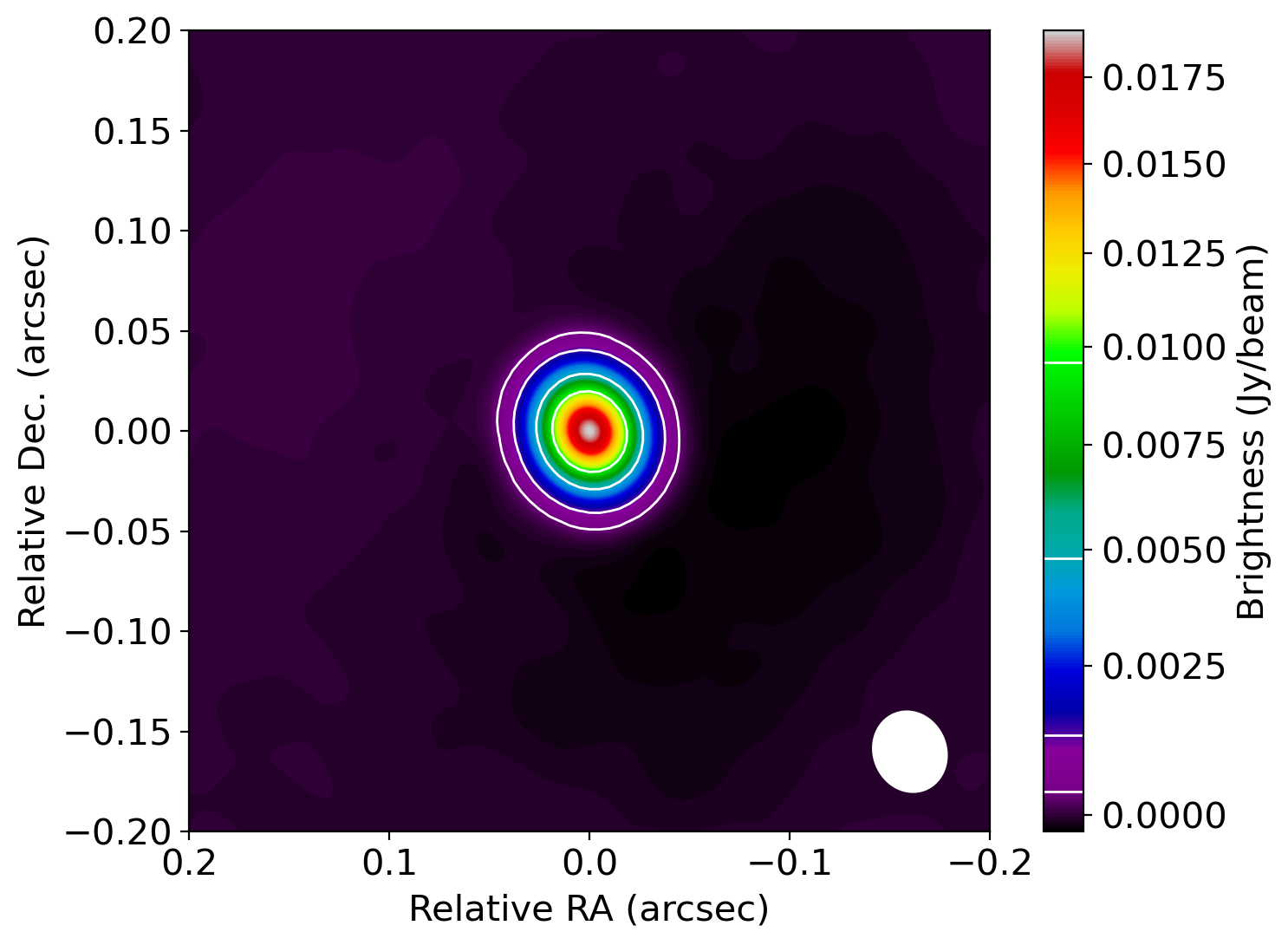}
	\caption{Continuum emission of the R Aql system. Contours are drawn at $3,12,48,$ and $96$ times the continuum rms noise value ($1 \times 10^{-4} \, {\rm Jy \, beam^{-1}}$).}
	\label{Fig:continRAql}
\end{figure}

In the observational images, $1 \, {\rm arcsec}$ corresponds to a spatial extent of $230 \, {\rm au}$.
The continuum emission of the combined \textsc{ATOMIUM} ALMA dataset is shown in Fig.~\ref{Fig:continRAql}.
It consists of one centrally condensed patch of emission with a specific intensity peak of $18.9 \, {\rm mJy \, beam^{-1}}$.
The total flux density within the $3 \times \sigma_{\rm rms}$ contour with a radius of $0.05 \, {\rm \arcsec}$, is $19.5 \,{\rm mJy \, beam^{-1}}$.
Except for a slight elongation of the emission patch due to the beam shape, no deviations from spherical symmetry are resolved, and no close companions can be identified in this way.
Outside the $3 \times \sigma_{\rm rms}$ contour, the emission is dominated by Gaussian noise with a mean close to zero, which is typical for an image obtained by interferometry.

\begin{figure*}
	\includegraphics[width = \textwidth]{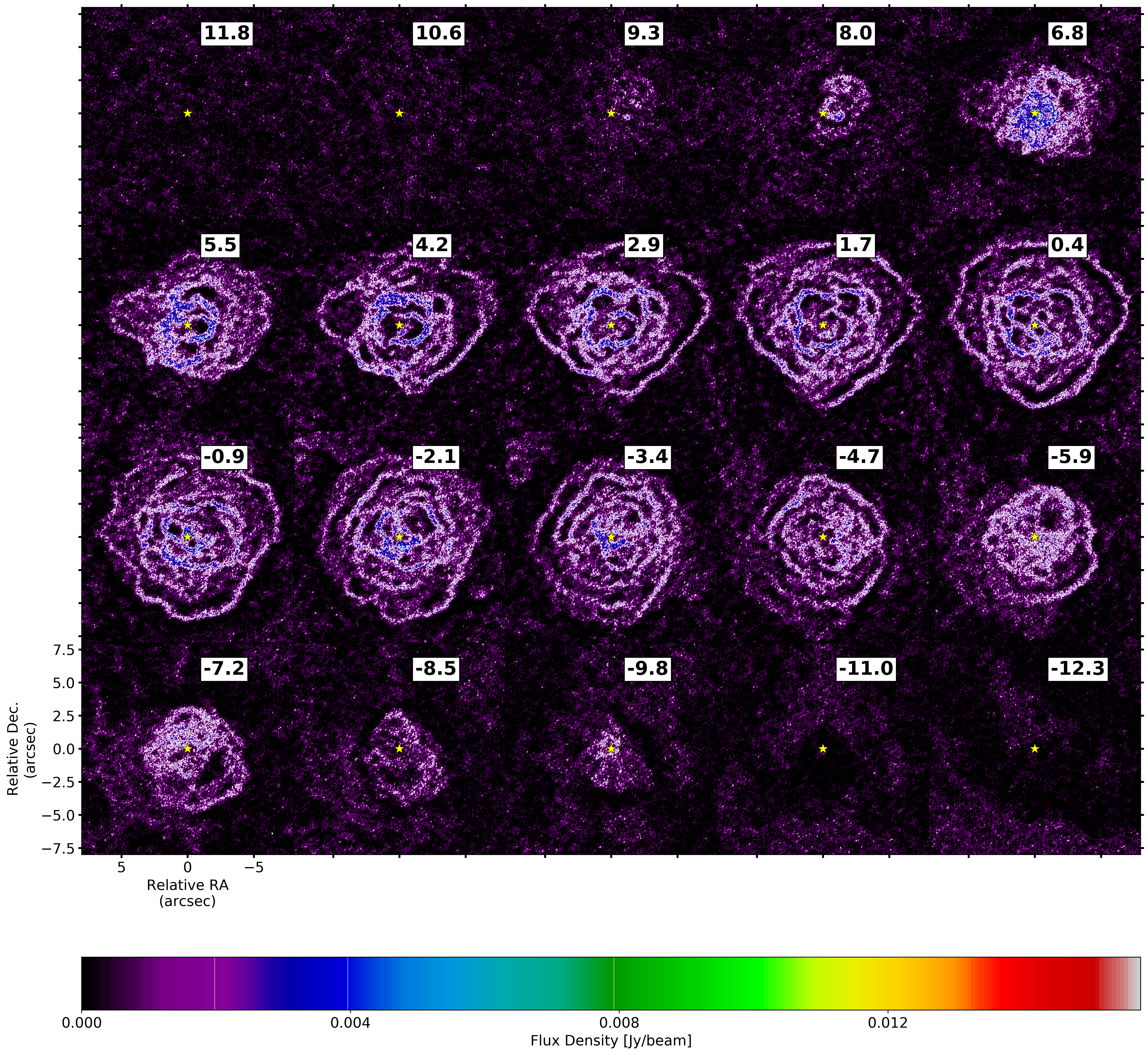}
	\caption{Channel maps of the combined $^{12}$CO $J=2$\textrightarrow$1$ emission of R Aql, with blue- and red-shifted projected velocities {(annotated in upper right corner in ${\rm km \, s^{-1}}$)}, ranging from $-12.3$ to $11.8 \, {\rm km \, s^{-1}}$ with respect to the stellar velocity of $47.2  \, {\rm km \, s^{-1}}$. Contours are drawn at $3,6$ and $12$ times the rms noise value of $6.6 \cdot 10 ^{-4}$ Jy/beam.}
	\label{Fig:RAql_CO}
\end{figure*}

Emission maps of the combined $^{12}$CO $J=2$\textrightarrow$1$ emission, with blue- and redshifted projected velocities ranging from $-12.3$ to $11.8 \, {\rm km \, s^{-1}}$ with respect to the stellar velocity of $47.2  \, {\rm km \, s^{-1}}$ are shown in Fig.~\ref{Fig:RAql_CO}. 
In most channels, the bulk of emission is confined within a roughly circular region centred around the star, so there is no clear sign of rotation or deviation from a roughly spherical, radial velocity profile. 
The emission has a maximal radial extent of about $14 \arcsec$ (corresponding to $3200 \, {\rm au}$) around the central velocity, and decreases for increasing projected velocity.
Around the maximal velocity of $\approx \pm 9$-$10 \, {\rm km \, s^{-1}}$, at the redshifted side ($8.0$ and $9.3 \, {\rm km \, s^{-1}}$ channel maps) the emission is located more to the north, whereas at the blueshifted side ($-8.5$ and $-9.8$ channel maps) there is an asymmetry towards the south. This indicates that the global morphology could be slightly oblate instead of spherical, and inclined with the side towards us pointing slightly to the south, corresponding to a small positive inclination $i$.
Further, at smaller scales, there are many intricate inhomogeneities including non-circular concentric shells, arcs and spiral-like structures.
For example, the most outer emission structure that extends up to $7 \arcsec$ at the central velocities, appears as a triangular morphology in red-shifted channels $v = 0.4$ to $v =4.2 \, {\rm{km \, s^{-1}}}$, and it is rather circular in the blue-shifted channels. At smaller radii, there are many more intricate structures that are hard to disentangle and identify as shells, arcs or spirals.

\subsection{Comparison to synthetic observation}
\label{ch:RAqlcomparison}

Now that we understand the general impact of a triple companion on the resulting wind structures of mass-losing stars from Sect.~\ref{ch:triples}, we compare these with the observations of R Aql. 
The typical features appearing in the simulations of triple systems, being an inner spiral (or small arcs) with an additional larger scale spiral (or larger scale arcs), and a waved pattern in the outer spiral, are not distinguishable in the observed $^{12}$CO $J=2$\textrightarrow$1$ emission of R Aql in Fig.~\ref{Fig:RAql_CO}. However, it is not straightforward to compare the 2D slices of our 3D simulations (in Sect.~\ref{ch:triples}) to these channel maps.
Firstly, the channel maps are not slices in the $x$-$y$-$z$ spatial dimensions, but 2D maps with material with a certain projected velocity towards the observer. To locate the material on these channel maps in the 3D space, one needs to know the {entire} wind velocity field, which is generally unknown and can be very complicated \citep[see e.g.][]{DeCeuster2024}. 
Secondly, the densities in the simulations do not represent exactly the observed $^{12}$CO $J=2$\textrightarrow$1$ emission.
Therefore, we need to post-process the output of our models with a radiative transfer code, to calculate the level populations and flux densities, and in this way provide synthetic channel maps and line profiles of certain observed molecules. Additionally, to mimic the process of the interferometric observations and the resulting noise and lost flux, an ALMA-simulator needs to be applied to these synthetic observables.
Thirdly, in the hydrodynamic simulations in Sect.~\ref{ch:triples}, the inclination has a big impact on how the different structures and global morphologies appear. In observations, the inclination {and orientation} at which the system is viewed is generally unknown, since the orbit and location of the possible companion(s) are unconstrained, but indications {for the inclination} can be found from analysing specific observed structures or global asymmetries. 
The observations of R Aql might have a slightly positive inclination and oblate global morphology, given the asymmetric location of the emission around the terminal projected velocities (see Sect.~\ref{ch:ObservRAql}).

We estimate with large uncertainty that the average radial distance between two observed arc- and spiral-like structures in the outer part of the outflow morphology in the central velocity channels in Fig.~\ref{Fig:RAql_CO} is $\sim 1.15 \, {\rm arcsec}$, indicating that these wind structures travel about $\sim 265 \, {\rm au}$ during one orbital period.
Assuming that the radial wind velocity in this region is $\sim 10 \, {\rm km \, s^{-1}}$, the time between two structures (corresponding to the orbital period), is $\sim 126 \, {\rm yrs}$.
Model v08m06 (with characteristics given in Tables~\ref{Ta:setupTableTriples} and ~\ref{Ta:inputTableTriples}) consists of an inner binary $M_1 + M_2 = 2 \, {\rm M_\odot}$ and outer companion $M_3 = 0.6 \, {\rm M_\odot}$ at a separation of $35 \, \rm au$ and initial wind velocity below $10 \, {\rm km \, s^{-1}}$. Given that its outer orbital period is $P_{\rm o}=128.5 \, {\rm yrs}$, this is a model that could represent the R Aql system.

We therefore compare the morphology in the observation and simulation of model v08m06, to investigate if the outflow of R Aql might be shaped by a similar triple system.
To do this, we post-process the \textsc{Phantom} output of model v08m06 with the \textsc{Magritte}\footnote{Open source \& available online: \href{https://github.com/Magritte-code/Magritte}{https://github.com/Magritte-code/Magritte}.} radiative transfer code, which is a ray-tracer that solves the radiative transfer equation using a second-order Feautrier scheme along each ray (for more details see \citealp{DeCeuster2020b,Deceuster2020a,DeCeuster2022}).
Since the hydro-simulations do not model the chemistry, we assume that all mass is in H$_2$, and the fractional abundance of CO is [CO$/$H$_2$]$= 3\times 10^{-4}$, which is a reasonable assumption for an O-rich AGB star \citep{VanDeSande2019}. \textsc{Magritte} uses the atomic and molecular database LAMDA \citep{LAMBDA2005}, and the calculations are done in non local thermodynamical equilibrium (NLTE), taking into account the first $40$ rotational transitions. The micro-turbulence is set to $0.5 \, {\rm km \, s^{-1}}$ \citep{Ramstedt2008}, and the cosmic microwave background radiation is taken as outer boundary condition.

\begin{figure*}
    \centering
    % \includesvg[width = \textwidth]{figures/v08m06_nLTE_i30_HR_notReduced.svg}
    % \includegraphics[width = \textwidth]{figures/v08m06_nLTE_i30_HR_notReduced_newColorbar.pdf}
    \includegraphics[width = \textwidth]{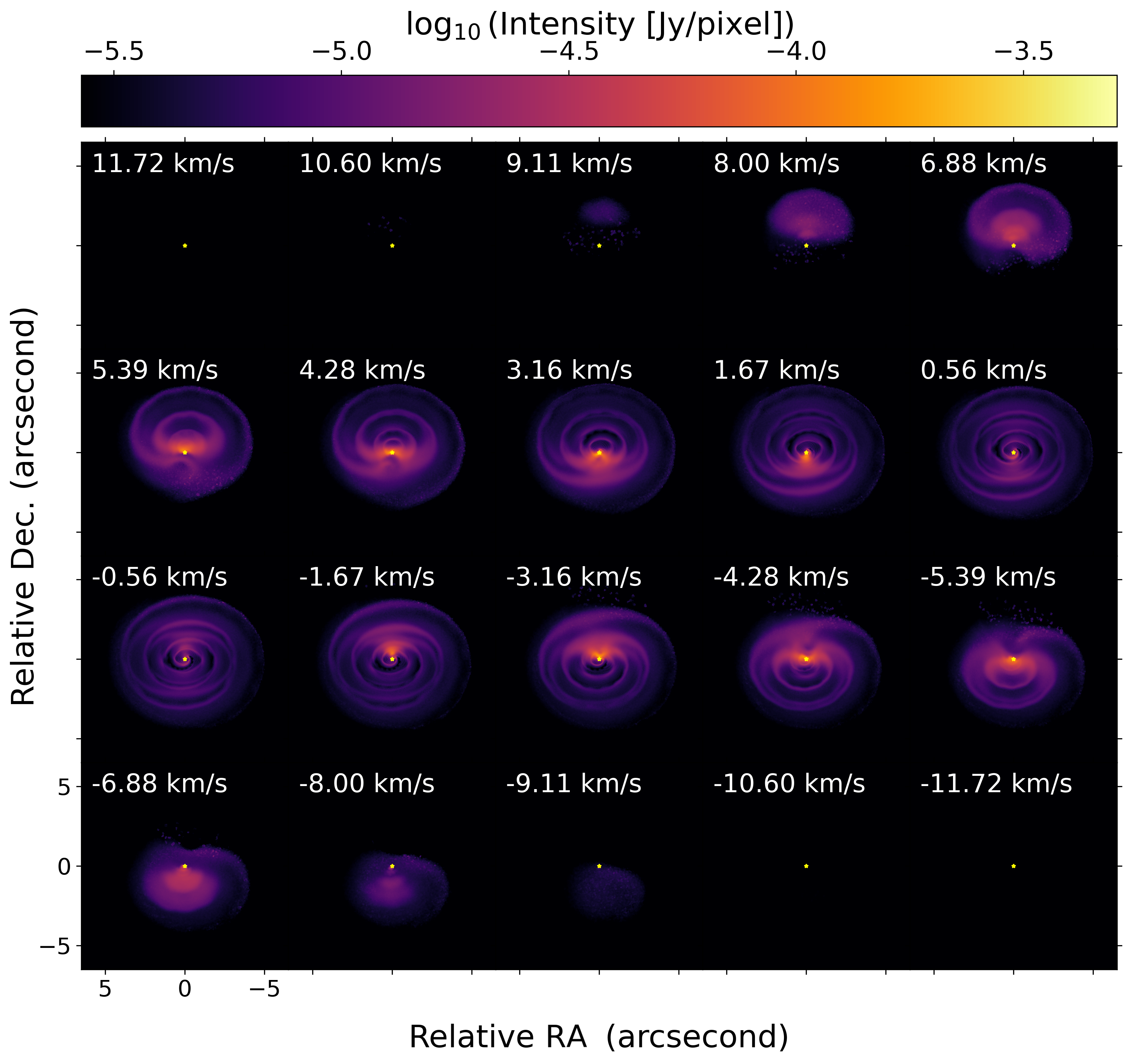}
    \caption{Synthetic channel maps of the $^{12}$CO $J=2$\textrightarrow$1$ emission of model v08m06, post-processed with the radiative transfer code \textsc{Magritte} \citep{DeCeuster2020b,Deceuster2020a,DeCeuster2022}, assuming an inclination of $30 \degree$ and distance of $230 \, {\rm pc}$. A yellow star is added indicating the location of the central star.}
    \label{fig:MagrChMaps_v08m06}
\end{figure*}

Assuming the system is observed at the same distance as R Aql ($230 \, {\rm pc}$) and under an inclination of $30 \degree$, the simulated $^{12}$CO $J=2$\textrightarrow$1$ intensity at similar projected velocities as in the R Aql channel maps (Fig.~\ref{Fig:RAql_CO}) is shown in Fig.~\ref{fig:MagrChMaps_v08m06}. 
The morphology consists of bright emission patches that move from north to south of the star between the projected velocities of $8$ to $0 \, {\rm km \, s^{-1}}$ and $0$ to $-8  \, {\rm km \, s^{-1}}$. This corresponds to wind material residing close to the polar regions, and producing a $2$-peak pattern on the emission line. 
The bright emission patches lie on top of the underlying large-scale spiral pattern (gravity wake of outer companion $M_3$), that is deformed compared to the orbital plane structure in the model, due to the inclination and the uncertain deprojected 3D coordinates of the wind material in each channel map. The underlying close binary spiral pattern, and the wave pattern in the outer spiral edge are more difficult to distinguish in these channel maps, that also ignore the noise simulations by ALMA. Fig.~\ref{fig:RAqlZoom} shows in more detail the central channel map with $v_{\rm proj} = 0.56 \, {\rm km \, s^{-1}}$. The compact spiral pattern is visible within the central region (within $\sim 1 \, {\rm arcsec}$), and the waved pattern of the outer edge of the $M_3$ spiral wake is distinguishable at small positive relative {RA}.
We additionally performed a test in which we post-process a simulation of the outer binary of this triple system, with $M_{\rm p} = 2.0 \, {\rm M_\odot}$, $M_{\rm s} = 0.6 \, {\rm M_\odot}$, $v_{\rm ini} = 8 \, {\rm km \, s^{-1}}$, $a = 35 \, {\rm au}$. Assuming the same inclination and distance as the triple model in Fig.~\ref{fig:MagrChMaps_v08m06}, we found that with the binary simulation, a clear spiral pattern and less asymmetry appear in the channel maps, making the two cases distinguishable.

Comparing the synthetic observations of the triple model (Fig.~\ref{fig:MagrChMaps_v08m06}) to the actual R Aql data (Fig.~\ref{Fig:RAql_CO}), the global morphology has a similar shape. It is quasi-spherical around the central velocity channels, and decreases in radial extent with increasing velocities. In both the actual and synthetic observations, at the maximal projected velocities, an emission patch is visible above and below the stars, which is due to the inclination and elongated morphology (see edge-on density profile of model v08m06 in Fig.~\ref{fig:tr_v08_rho_z1130}). 
When looking at the smaller-scale structures, in both the observation and simulations, there are rather intricate spiral-like structures. A {detailed} one-to-one comparison is, however, not feasible, due to the large uncertainty on the selected parameters of the model and the inclination, and because the SPH models lack some physics such as the impact of the stellar pulsations, that may affect the wind morphology. {One has to be very careful with interpreting any more detailed (non-global) resemblances between the model and observation. Given the before mentioned various uncertainties, in this case, such resemblances may be rather coincidental and be created by entirely different physical processes.}
Further, a more direct comparison of the \textsc{Magritte} synthetic observables to the observations requires to post-process the data with an ALMA-simulator, to model the noise. This is beyond the scope of this work, as it would further complicate the comparison.
R Aql might be shaped by a similar triple system, under a slight positive inclination, but further detailed studies on the observational data are required to discover the actual nature of this system.

\section{Conclusion}
\label{ch:conclusions}
With the use of high-resolution 3D Smoothed-Particle-Hydrodynamic (SPH) \textsc{Phantom} models, we investigated the effect of a co-planar hierarchical triple system on the morphology of AGB winds.

In general, adding an outer companion $M_3$ to an inner binary $M_1$--$M_2$ results in an additional, large-scale spiral structure, on top of the more compact spiral structure created by the inner binary. This additional large-scale spiral is a gravity wake behind the outer companion, and has a waved outer edge, due to its interaction with the underlying compact $M_1$--$M_2$ spiral structure. 
The added companion $M_3$ also induces an orbital motion of the inner binary $M_1$--$M_2$ around the system's centre-of-mass. This makes the inner $M_1$--$M_2$ spiral deviate from an Archimedes spiral, and resemble a snail-shell pattern.
When decreasing the wind velocity, we find that, similar to the binary case, the global morphology significantly changes and becomes more flattened, and the impact of the companions on the morphology increases. 

We also explored the impact of eccentric orbits on these morphologies. We find that the effect of introducing eccentricity in the inner orbit on the large-scale structures is limited. On smaller scales, this introduces asymmetries characteristic for eccentric systems, as was studied in \cite{Malfait2021}.
If the outer orbit is eccentric, the effect is larger, but still less pronounced then in the binary model, indicating that the triple nature reduces the effect of eccentricity. The signatures of eccentricity are clearest when comparing the large-scale edge-on structures at apastron and periastron side of the companion's orbit. 
In an eccentric triple system, very complicated, phase-dependent wind-companion interactions take place, resulting in asymmetric spiral and arc structures. 

Finally, we included a brief study of the wind structures emerging in the complex outflow of the AGB star R Aql, observed within the \textsc{ATOMIUM} ALMA large programme. Due to its chaotic morphology, this star could host a triple system. Although a detailed one-to-one comparison with one of our simulations is unfeasible, we use he radiative transfer code \textsc{Magritte} to post-process model v08m06, which could resemble the R Aql system observed at an inclination of $\sim 30 \degree$. From the modelled and observed $^{12}$CO $J=2$\textrightarrow$1$ emission, it is not straightforward to identify the triple nature of the system from the emission maps, because the characteristics identified in the triple models are not resolved in the channel maps. 
The morphology of R Aql could be the result of a similar triple system as the ones studied in this work, but no clear direct implications of this are detected, and a more detailed study of the observational data of R Aql is required. 

A direct comparison between observations and simulations of AGB outflows is very challenging, due to the large uncertainties in the wind and orbital properties, and the uncertainties that come with comparing a 3D simulation to 2D projected velocity maps of a wind with uncertain velocity field.
Additionally, more relevant physical, chemical \citep{Maes2024}, and approximate radiation \citep{Esseldeurs2023} processes, which may play a role in shaping the observed outflow, are still being added to the SPH wind model.
Alternatively, in contrast to the bottom-up approach presented in this paper, model reconstruction methods might help to directly obtain probable 3D models for our most complex observations \citep[see e.g.][]{DeCeuster2024}.

This work provides an important first step in modelling the wind structures resulting from the interaction with companions beyond the binary paradigm.

%### Acknowledgements
%###################################################

\begin{acknowledgements}

The computational resources and services used in this work were provided by the VSC (Flemish Supercomputer Center), funded by the Research Foundation Flanders (FWO) and the Flemish Government – department EWI.
We thank Daniel Price for his help, especially concerning the development of \textsc{Phantom}. The animations and figures in this paper are made using the visualisation tools \textsc{Plons}\footnote{\url{https://github.com/Ensor-code/plons}} and \textsc{Splash} \citep{splash}.
ALMA is a partnership of ESO (representing its member states), NSF (USA) and NINS (Japan), together with NRC (Canada) and NSC and ASIAA (Taiwan), in cooperation with the Republic of Chile. The Joint ALMA Observatory is operated by ESO, AUI/NRAO and NAOJ.
JM, ME, and LD acknowledge support from the KU Leuven C1 excellence grant C16/17/007 MAESTRO and the FWO research project G099720N. L.S. is a senior F.N.R.S research associate.
 FDC is a postdoctoral research fellow of the Research Foundation - Flanders (FWO), grant 1253223N.
LD acknowledges support from the KU Leuven C1 excellence grant BRAVE C16/23/009 and KU Leuven Methusalem grant SOUL METH/24/012. SHJW acknowledges support from the Research Foundation Flanders (FWO) through grant 1285221N.

\end{acknowledgements}

%### Bibliography
%###################################################

\bibliographystyle{aa}
\bibliography{references}

\appendix

\section{Additional figures hydrodynamic simulations}
This section contains additional figures of the wind morphology in binary and hierarchical triple systems.

\begin{figure*}[h!]
    \centering
    \includegraphics[width = 0.4\textwidth]{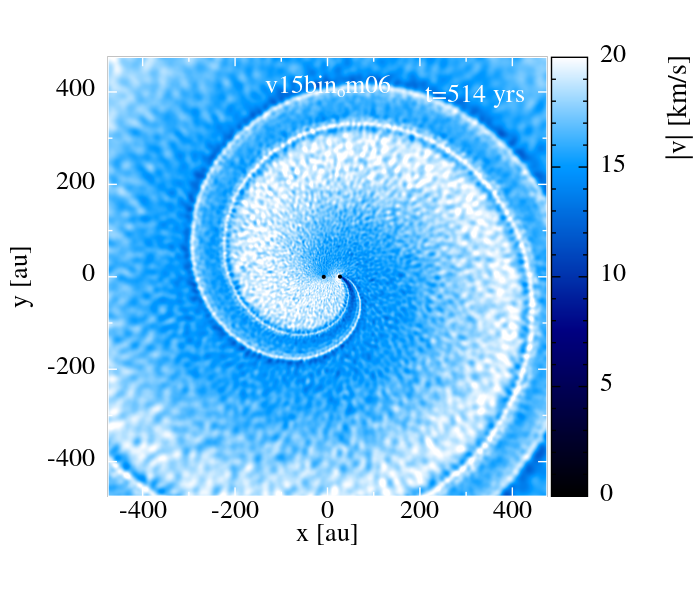}
    \caption{Velocity distribution in a slice through the orbital plane of model v15bin$_{\rm o}$m06 for $r<475 \, {\rm au}$.}
    \label{fig:wideBin_vel_z475}
\end{figure*}
\begin{figure*}
    \centering
    \includegraphics[width = 0.7 \textwidth]{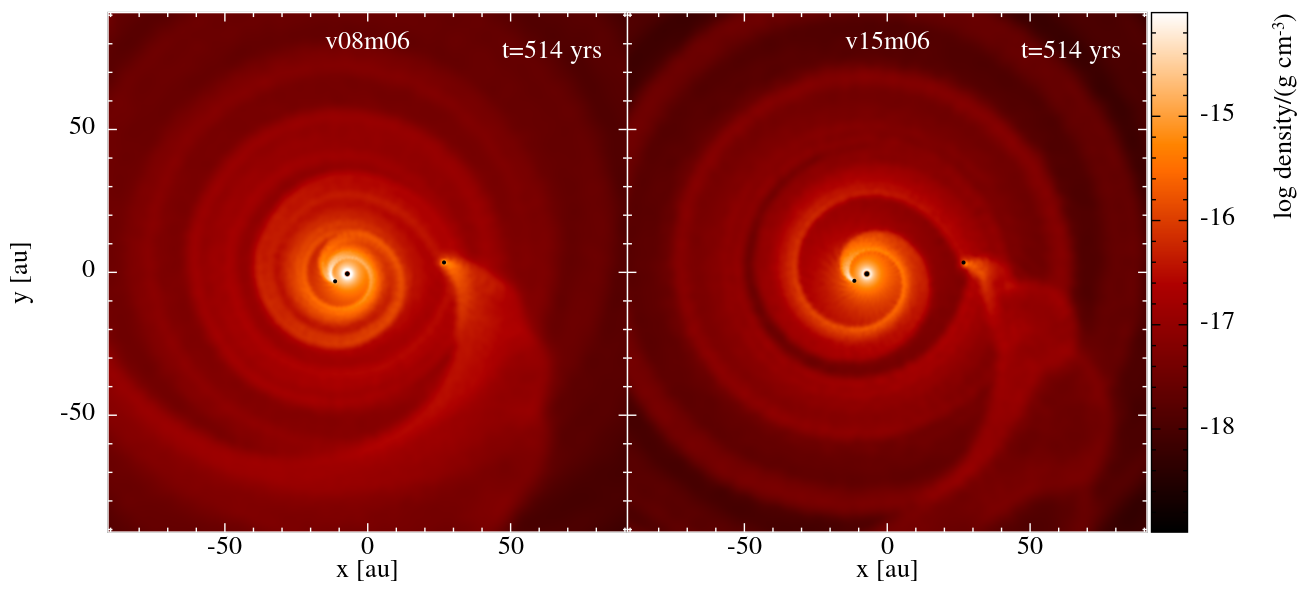}
    \caption{Density distribution in a slice through the orbital plane of models v08m06 (left) and v15m06 (right) for $r<90 \, {\rm au}$. }
    \label{fig:v08_v15_m06_z100}
\end{figure*}

\section{Additional figure synthetic observation}
This section contains an additional figure of the synthetic observation of model v08m06.

\begin{figure*}
    \centering
    % \includesvg[width = 0.7 \textwidth]{figures/v08m06_nLTE_i30_zoomed_HR_notReduced.svg}
    % \includegraphics[width = 0.7 \textwidth]{figures/v08m06_nLTE_i30_zoomed_HR_notReduced_newColorbar.pdf}
    \includegraphics[width = 0.7 \textwidth]{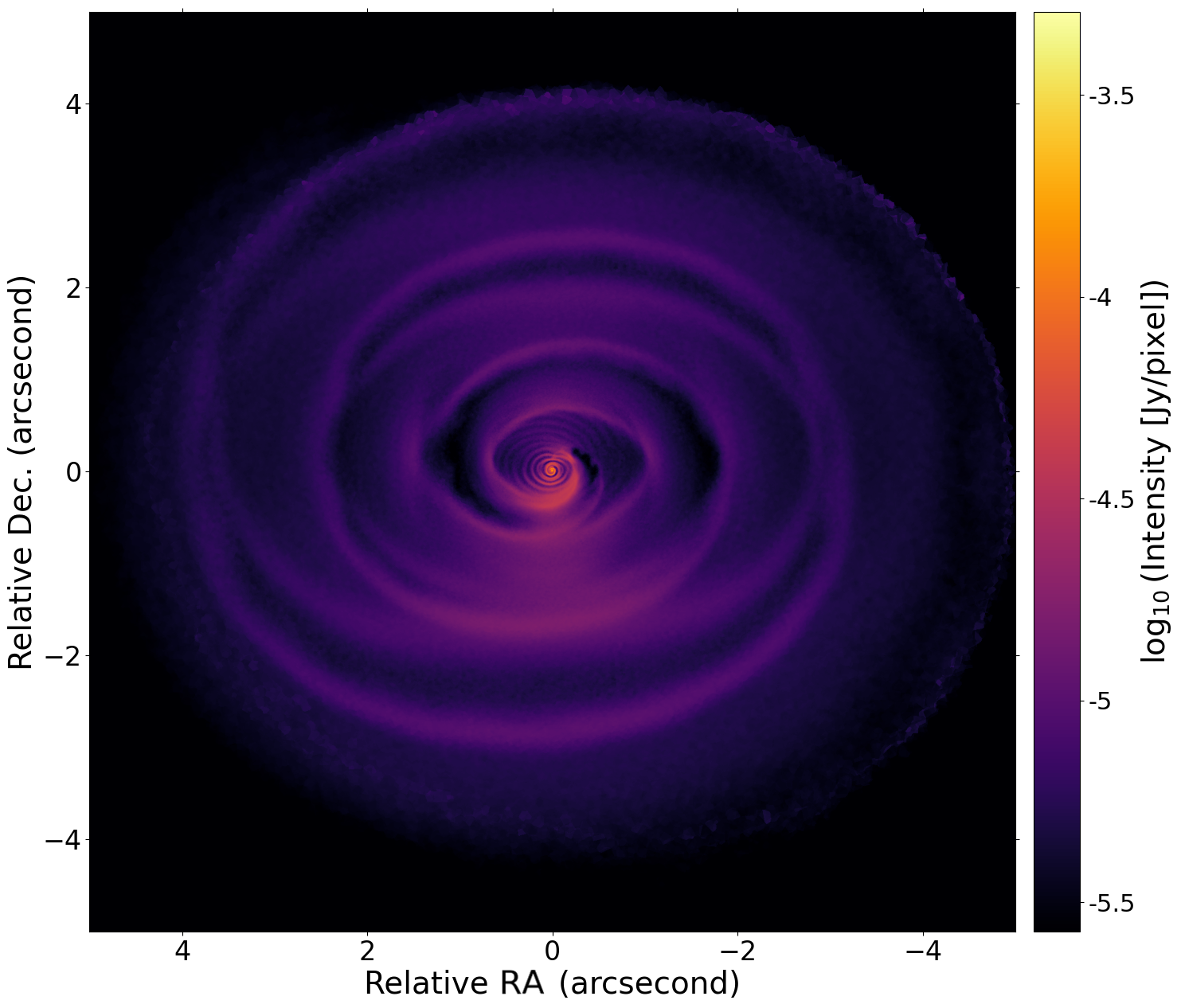}
    \caption{Synthetic channel map of the $^{12}$CO $J=2$\textrightarrow$1$ emission with $v=0.56 \, {\rm km \, s^{-1}}$ of model v08m06 presented in Fig.~\ref{fig:MagrChMaps_v08m06}.}
    \label{fig:RAqlZoom}
\end{figure*}

\end{document}